\documentclass{aa}  

\usepackage{graphicx}
\usepackage{txfonts}
\usepackage{hyperref}
\usepackage{color}
\usepackage{placeins}
\newcommand\lsim{\mathrel{\rlap{\lower4pt\hbox{\hskip1pt$\sim$}}
    \raise1pt\hbox{$<$}}}
\newcommand\gsim{\mathrel{\rlap{\lower4pt\hbox{\hskip1pt$\sim$}}
    \raise1pt\hbox{$>$}}}

\newcommand{\fwidth}{0.9\hsize}
\newcommand{\fheight}{0.55\hsize}
\renewcommand{\S}{Sect.}

\begin{document}

   \title{The S stars' zone of avoidance in the Galactic center}


   \author{Aleksey Generozov
          \inst{1}\fnmsep\thanks{Corresponding author: aleksey.generozov@gmail.com}
          \and
          Hagai B. Perets\inst{1,2}
          \and
          Matteo S. Bordoni\inst{3}
          \and
          Guillaume Bourdarot\inst{3}
          \and
          Antonia Drescher\inst{3}
          \and
          Frank Eisenhauer\inst{3,4}
          \and
          Reinhard Genzel\inst{3,5}
          \and 
          Stefan Gillessen\inst{3}
          \and
          Felix Mang\inst{3}
          \and
          Thomas Ott\inst{3}
          \and
          Diogo C. Ribeiro\inst{3}
          \and
          Rainer Sch{\"o}del\inst{6}
          }

        \institute{Physics Department, Technion - Israel Institute of Technology, 3200003
        Haifa, Israel
         \and
         Astrophysics Research Center of the Open University (ARCO),
         4353701 Raanana, Israel
         \and
        Max-Planck-Institut für Extraterrestrische Physik,
        D-85748 Garching, Germany
         \and 
         Department of Physics, Technical University of Munich,
         D-85748 Garching, Germany
         \and
         Departments of Physics \& Astronomy, University of California,
         Berkeley, CA 94720, USA
         \and
        Instituto de Astrofísica de Andalucía,
        18008 Granada, Spain
        }

   \date{November 24, 2024}
   \abstract{
   This paper investigates the origin and orbital evolution of S stars in the Galactic center using models of binary disruption and relaxation processes. We focus on explaining the recently discovered ``zone of avoidance'' in S-star orbital parameters, defined as a region where no S stars are observed with pericenters of $\log(r_p / {\rm AU}) \leq 1.57 + 2.6(1 - e)$ pc. We demonstrate that the observed S-star orbital distributions, including this zone of avoidance and their thermal eccentricity distribution, can be largely explained by the continuous disruption of binaries near the central supermassive black hole, followed by orbital relaxation. Our models consider binaries originating from large scales ($5$--$100$ pc) and incorporate empirical distributions of binary properties. We simulate close encounters between binaries and the black hole, tracking the remnant stars' orbits. The initially highly eccentric orbits of disrupted binary remnants evolve due to nonresonant and resonant relaxation in the Galactic center potential. While our results provide insights into the formation mechanism of S stars, there are limitations, such as uncertainties in the initial binary population and mass function and simplifications in our relaxation models. Despite these caveats, our study demonstrates the power of using S-star distributions to probe the dynamical history and environment of the central parsec of our Galaxy.}
    \titlerunning{The zone of avoidance in the Galactic center}
    \authorrunning{Generozov et al.}
    \maketitle

%

\section{Introduction}
The Galactic center's proximity allows for the direct measurement of stellar orbits within the central parsec. There are distinct structures within this region: multiple young stellar disks with O and Wolf--Rayet stars \citep{levin&beloborodov2003,paumard+2006,bartko+2009,yelda+2014,vonFellenberg+2022}, as well as the isotropic S-star cluster with many B-and later-type stars \citep{gillessen+2017}.

These stars have been the subject of intense, decades-long studies. They have established the presence of the central supermassive black hole (SMBH) in the Galactic center \citep{genzel+1996,genzel+1997,ghez+1998}, and the closest S stars can be used for stringent tests of general relativity \citep{gravity+2018_redshift,do+2019,gravity+2020_S2prec}. The S stars can also be used to probe the extended mass distribution in the Galactic center \citep{gravity+2020_S2prec,gravity+2022_mass}. 

The origin of the S stars is a long-standing puzzle, considering that strong tides from Sgr A* are expected to suppress star formation in the S-star region \citep{sanders1992, ghez+2003}. Instead, the S stars may have migrated inward from larger scales via tidal disruption of binary stars \citep{hills1988,Gou+03}. Binaries may either come from relatively large scales ($\sim$ tens of parsecs; \citealp{perets+2007,Ham+17}) or from one of the young disks \citep{madigan+2009,generozov&madigan2020,rantala&naab2023}.
Remnants from binary stars would initially be on highly eccentric ($e \gsim 0.98$; \citealp{hills1988}) orbits. In contrast, the observed S stars have a thermal eccentricity distribution. However, other stars or remnants can perturb the stars' orbits following binary disruptions. The stars' energy only evolves due to uncorrelated two-body encounters (nonresonant relaxation). The star's angular momentum can evolve on much faster timescales due to coherent torques (resonant relaxation; \citealp{rauch&tremaine1996, hopman&alexander2006, perets+2009, kocsis&tremaine2011, antonini&merritt2013}). The S stars can easily isotropize within their lifetime via vector resonant relaxation \citep{hopman&alexander2006}, but the eccentricity distribution relaxes on the slower scalar resonant relaxation time. Many papers have studied whether this process can reproduce the observed eccentricities within the stars' lifetimes.

For example, \citet{merritt+2009} and \citet{perets+2009} used N-body simulations to provide the first detailed theoretical study of the S stars' eccentricity evolution due to an intermediate-mass black hole (IMBH) and stellar-mass black holes, respectively. \citet{antonini&merritt2013} were the first to study resonant relaxation in the Galactic center with both relativistic and Newtonian precession. Initial studies used approximate treatments for resonant relaxation. Subsequently, a self-consistent formalism was developed \citep{bar-or&alexander2014, sridhar&touma2016, fouvry+2017, bar-or.fouvry2018} and applied to the S stars by \citet{generozov&madigan2020} and \citet{tep+2021}.

Generally, these studies find that a population of stellar-mass black holes is needed to reproduce the S stars' eccentricity within their lifetimes. The exact number and mass required depend on the modeling assumptions, the observational dataset, and the uncertain ages of the S stars.
The other orbital elements provide complementary constraints on the S stars' origin. For example, \citet{perets.gualandris2010} found that many theoretical models have difficulty explaining the populations of B stars on small and large scales.

We revisit the problem of the S stars' orbital evolution, in light of the most recent S-star observations and constraints on the background mass distribution. For example, recently it was shown that there is a dearth of S stars with pericenters of

\begin{equation} 
\log(r_p / {\rm AU}) \leq 1.57 + 2.6 \left(1 - e\right)
\label{eq:za}
,\end{equation}
even though they would be observable \citep{burkert+2023}. The S stars' eccentricity distribution in the latest GRAVITY data is thermal, within observational uncertainties. The zone of avoidance constrains the semimajor axis distribution of the progenitor binaries, which depends on their intrinsic distribution and the rate of loss cone refilling. The eccentricity distribution of the S stars is sensitive to the properties of the background population, including the mass distribution of stellar-mass black holes.

 We show that these orbital distributions can be explained via a combination of binary disruptions and relaxation in the Galactic center. If binaries are continuously disrupted in the Galactic center, the observed eccentricity distribution can be reproduced with a population of $\sim10 M_{\odot}$ black holes in the S-star cluster. In the case of impulsive injection, as is expected in the young disk scenario, the background black holes would have to be at least $\sim 30 M_{\odot}$, considering existing constraints on the total enclosed mass in the Galactic center. 

Our model accounts for many processes that to our knowledge have never been studied together in the literature. In particular, we have (i) an observationally motivated progenitor binary population, (ii) self-consistent diffusion coefficients for resonant relaxation, (iii) both eccentricity and semimajor axis evolution, (iv) stellar evolution, and (v) a loss cone. We also model the K-band luminosity function (KLF) of the S stars and incorporate this into our observational comparisons.   

The remainder of this paper is organized as follows. In \S~\ref{sec:methods},
we discuss our methods and the relevant physical processes. In \S~\ref{sec:res},
we show that most S-star observables can be reproduced by continuous disruption of binaries in the Galactic center. In \S~\ref{sec:add}, we discuss additional effects that are not accounted for in our fiducial models.  In \S~\ref{sec:imp}, we discuss models where binaries are all disrupted at a fixed lookback time. We summarize our main results in \S~\ref{sec:conc}.

\section{Relevant physical processes and methods}
\label{sec:methods}
\subsection{Binary disruptions}
\label{sec:bprop}
We assume that all the S stars are initially sourced by the disruption of binaries that come near their tidal radius:

\begin{align}
    r_t &= \left(\frac{M_{\rm bh}}{m_{\rm bin}}\right)^{1/3} a_{\rm bin}\nonumber\\
        &\approx 3.7 \times 10^{-4}\, {\rm pc} \left(\frac{M_{\rm bh}}{4.3 \times 10^6 M_{\odot}}\right)^{1/3} \left(\frac{m_{\rm bin}}{10 M_{\odot}}\right)^{-1/3} \left(\frac{a_{\rm bin}}{1 {\rm au}}\right),
    \label{eq:rt}
\end{align}
where $M_{\rm bh}$ and $m_{\rm bin}$ are the mass of the black hole and binary, and $a_{\rm bin}$ is the binary semimajor axis.
During a tidal interaction with an SMBH, a binary can (i) collide,
(ii) survive (with some change to the orbital elements), or (iii) 
disrupt \citep{hills1991, ginsburg&loeb2007,antonini+2010,bradnick+2017}. In the latter case, the binary is split into two stars. In the limit
of a parabolic center-of-mass orbit, one of the stars is always bound to the SMBH, while the other is unbound. The semimajor axis of the remnant star is (e.g., \citealp{sari+2010})

\begin{align}
    a_s  &= \chi \left(\frac{M_{\rm BH}}{m_{\rm bin}}\right)^{2/3} \left(\frac{m_{\rm bin}}{m_{\rm ej}}\right) a_{\rm bin},\nonumber\\
    &\approx 0.028\, {\rm pc} \left(\frac{M_{\rm bh}}{4.3 \times 10^6 M_{\odot}}\right)^{2/3} \left(\frac{m_{\rm bin}}{10 M_{\odot}}\right)^{-2/3} \left(\frac{m_{\rm bin}}{m_{\rm ej}}\right) \left(\frac{a_{\rm bin}}{1 {\rm au}}\right) ,
    \label{eq:as}
\end{align}
where $m_{\rm ej}$ is the mass of the ejected star and $\chi$ is typically a factor of order unity. The pericenter of the remnant star, $r_{p,s}$, is 
comparable to that of the progenitor binary. Thus, $r_{p, s} \lsim r_t$.
The eccentricity of the remnant star, $e_s$, is close to one.
More precisely, 
\begin{align}
e_s &\gsim 1 - \left(\frac{m_{\rm ej}}{m_{\rm bin}}\right) \left(\frac{M_{\rm BH}}{m_{\rm bin}}\right)^{-1/3} \nonumber \\
     &\gsim 1 - 0.013 \left(\frac{m_{\rm ej}}{m_{\rm bin}}\right)  \left(\frac{M_{\rm bh}}{4.3 \times 10^6 M_{\odot}}\right)^{-1/3} \left(\frac{m_{\rm bin}}{10 M_{\odot}}\right)^{1/3}  
     \label{eq:es}
.\end{align}
Binaries can have a few different sources: the clockwise disk \citep{madigan+2009,generozov&madigan2020,rantala&naab2023}, scattering from massive perturbers
such as large-scale molecular clouds or stellar clusters $\sim5- 100$ pc away from the Galactic center \citep{perets+2007}, or large-scale gaseous spiral arms at similar scales \cite{Ham+17}. We focus on the massive perturbers' origin here.

To model binary disruptions in the Galactic center, we first constructed a model field-like binary population such that 

\begin{enumerate}
\item The primary mass distribution follows a Kroupa mass function ($dN/dm_* \propto m_*^{-2.3}$ for $m_* >0.5 M_{\odot}$) \citep{kroupa2001}.
\item The binary properties (semimajor axis, eccentricity, and mass ratio) follow the empirical distributions in \citet{moe&distefano2017}. We only 
considered mass ratios above 0.1, considering the distribution is unconstrained
for smaller mass ratios. The higher multiplicity fraction of massive 
stars is accounted for.
\end{enumerate}

In principle, dynamical processes such as binary evaporation can affect the 
binary distribution. In practice, this is not relevant for binaries 
that would populate the S-star region. Specifically, binaries would disrupt 
prior to evaporating or hardening (see Appendix~\ref{sec:evap}). 

For each binary, we simulated a close encounter with a pericenter drawn from a uniform distribution (up to three times the maximum tidal radius of the binary population), implicitly assuming binaries are in the full loss cone regime as might be expected for the massive perturbers scenario. Furthermore, the S stars imply high rates of disruptions that are difficult to reproduce in the empty loss-cone regime \citep{perets+2007,perets.gualandris2010}. If the pericenter of the encounter is greater than three times the binary tidal radius, the binary is unlikely to be disrupted (see \citealp{sari+2010} and Fig.~5 in \citealp{generozov&madigan2020}), and thus unlikely to deposit stars in the S-star region. Such binaries were excluded from further analysis. Otherwise, we simulated a close encounter between the binary and the central SMBH with the \textsc{Fewbody} code \citep{fregeau+2004}.

For simplicity, we assumed a constant disruption rate. Model binaries were disrupted at a 
random look-back time between $10^9$ years ago and the present day. 
Furthermore, the binary was disrupted at a random point along the main sequence 
of the primary star. We discuss the implications of a disruption rate rising toward the present day in \S~\ref{sec:add}, motivated by recent observational results on the star formation history in the Galactic center. Such models may be able to better reproduce the age distribution and the KLF of the S stars. However, more massive background black holes may be required to reproduce the S stars' eccentricity distribution. Finally, we collected the remnants of all disrupted binaries with semimajor axes $<0.05$ pc, as S-star progenitors, and modeled their subsequent orbital relaxation.

\subsection{Relaxation}
\label{sec:rlx}
Binary disruptions deposit stars onto highly eccentric orbits (see equation~\ref{eq:es}), and 
cannot account for the nearly thermal S-star eccentricity distribution \cite{perets+2009}.
However, the orbits of remnant stars can evolve due to nonresonant and resonant relaxation \citep{hopman&alexander2006,perets+2009}. In the former, stars change their energy and angular momenta due to uncorrelated two-body encounters. In the latter, stars change their angular momenta due to coherent torques in potentials with a high degree of symmetry (like the nearly Keplerian potential in the Galactic center).

The nonresonant relaxation time at Galactocentric radius $r$ is 
\begin{align}
t_{\rm NRR} &\approx \frac{Q^2 P(r)}{N_*(r) \log(Q)}\nonumber\\
&\approx 5 \times 10^{10} {\rm\, yr} \left(\frac{m_*}{1 M_{\odot}}\right)^{-2} \left(\frac{N_*(1 \,\,{\rm pc})}{10^6}\right)^{-1} \left(\frac{r}{1 {\rm\, pc}}\right)^{\gamma- 3/2},
\end{align}
where $P(r)$ is the orbital period, $N_*(r)$ is the number of stars enclosed within $r$, $Q$ is the ratio of the SMBH mass to a characteristic stellar mass, and $\gamma$ is the power law index of the 3D density profile. It should be noted that the presence 
of stellar-mass black holes or other massive objects can significantly reduce the relaxation timescale (see review by \citealp{alexander2017}).

In the limit in which the precession is dominated by extended stellar mass, the resonant relaxation time is 
\begin{align}
t_{\rm RR} &\approx Q P(r) \nonumber \\
& \approx 1.5\times 10^{10} {\rm\, yr} \left(\frac{m_*}{1 M_{\odot}}\right)^{-1} \left(\frac{r}{1 {\rm\, pc}}\right)^{3/2}.
\end{align}
For highly eccentric orbits, general relativistic precession dominates and resonant relaxation is suppressed \citep{merritt+2011}.

These timescales depend on the background density profile of stars and remnants. 
We modeled the background with two components: (i) low-mass stars and remnants ($\sim 1 M_{\odot}$), and (ii) stellar-mass black holes ($\sim 10 M_{\odot}$). This is a reasonable approximation for modeling relaxation in an evolved galactic nucleus, and has been used extensively in the literature \citep{merritt2013}. In principle, the high-eccentricity S stars should also contribute to the background density. In practice, we assume that the enclosed mass is dominated by the older, relaxed population that dominates the central 10''
\citep{schodel+2020}.

Well inside the sphere of influence, a relaxed stellar profile will be between $r^{-1.5}$ and $r^{-1.75}$, depending on the relative abundance of stellar-mass black holes \citep{bahcall&wolf1976,alexander&hopman2009}. The black hole profile will be steeper than $r^{-2}$ on larger scales, and flatten to $r^{-1.75}$ in the innermost parts of the Galactic center \citep{freitag+2006,hopman&alexander2006a,alexander&hopman2009,preto&amaro-seoane2010,aharon&perets2015,vasiliev2017,zhang&amaro-seoane2024}.\footnote{For radii of $r\gsim 0.1$ pc, the BH profile can be somewhat flatter (e.g., \citealp{baumgardt+2018} find $r^{-1.55}$ in N-body simulations). However, this is outside the S-star cluster.} We used the following power law approximations for the density profiles of stars and black holes:

\begin{align}
&\rho_{\rm bh} = 7.2 \times 10^5  \left(\frac{r}{0.1 \,{\rm pc}}\right)^{-2.04} M_{\odot}\, {\rm pc}^{-3} \nonumber\\
&\rho_{\rm *} =  3.0 \times 10^6 \left(\frac{r}{0.1 \,{\rm pc}}\right)^{-1.46} M_{\odot}\, {\rm pc}^{-3}.
\label{eq:background}
\end{align}
This is a reasonable approximation for different Galactic center models.
Equation~\eqref{eq:background} is within $\sim 10\%$ of the density profiles from the Fokker-Planck 
models \citet{vasiliev2017} and within $\sim 40\%$ of the `Fiducial' model in
\citet{generozov+2018} between 0.01 and 0.1 pc after 10 Gyr of evolution. The final density
profiles of these models are similar, despite differences in the initial conditions. In \citet{vasiliev2017}, the stars and black holes all form together in the distant past. In \citet{generozov+2018}, the black holes
are completely sourced by ongoing star formation at the present-day location 
of the clockwise disk. Recent measurements of the diffuse starlight and resolved stars in the Galactic center indicate a stellar density
profile between $r^{-1.1}$ and $r^{-1.4}$ \citep{schodel+2018, gallego-cano+2018}. There is some evidence for stellar-mass black holes in the central parsec from observations of X-ray transients in this region \citep{mori+2019}. There is also a population of quiescent black hole X-ray binary candidates (\citealp{hailey+2018,mori+2021}, though the identification of these sources as black holes
is controversial \citealp{maccarone+2022}).

In this model, the total mass enclosed within the apocenter of S2 is $\sim 1.5\times 10^3 M_{\odot}$, comparable to the 1-$\sigma$ upper limit from the GRAVITY collaboration ($\sim 1000 M_{\odot}$; \citealp{gravity+2024extendedmass}). 
We note that, in principle, the present-day upper limit does not rule out a larger number of black holes in the past.

We calculated angular momentum diffusion coefficients for this model, using JuDOKA \citep{tep+2021}. Then we evolved the remnants' orbital elements forward in time, using the Monte Carlo procedure described in \citet{bar-or&alexander2016}. For every timestep, $\Delta t$, the reduced angular momentum, $j$,\footnote{The angular momentum divided by the circular angular momentum.} and the semimajor axis, $a$, are incremented by
\begin{align}
    &\Delta j = D_{\rm j,NRR} \Delta t + \gamma_1 \sqrt{D_{\rm jj,NRR} \Delta t} +
    D_{\rm j,RR} \Delta t \\
    &+ \gamma_3 \sqrt{D_{\rm jj,RR} \Delta t} \nonumber\\
    &\Delta a = D_{\rm a, NR} \Delta t + \gamma_2'(\gamma_1, \gamma_2) D_{\rm aa, NRR},
\end{align}
respectively, where $D_j$, $D_{jj}$, $D_a$, and $D_{aa}$ are the diffusion coefficients, and 
the subscripts ${\rm RR}$ and ${\rm NRR}$ denote resonant and nonresonant relaxation, respectively.  The unprimed $\gamma$ coefficients were independently drawn from a Gaussian distribution of unit variance, and
\begin{align}
&\gamma_2'(\gamma_1, \gamma_1) = \rho \gamma_1 + \sqrt{1-\rho^2} \gamma_2\nonumber \\
&\rho = \frac{D_{aj,NR}}{\sqrt{|D_{jj,NR} D_{a,NR}|}},
\end{align}
accounting for the covariance between energy and angular momentum diffusion.
We used an adaptive timestep for the evolution:
\begin{equation}
\Delta t = 10^{-3} \frac{j^2}{\mathrm{max}(D_{jj,RR}(j))},
\end{equation}
where $\mathrm{max}(D_{jj,RR}(j))$ is the maximum of $D_{jj,RR}$ at fixed energy. Gravitational wave emission was included, following \citet{peters1964}, but was negligible except for a small minority of stars (e.g., for 99\% of stars, the semimajor axis changes by less than 1\% with the semimajor axis diffusion artificially turned off). To speed up our calculations, we computed the diffusion coefficients for a predefined grid of semimajor axes and angular momenta. 

At each step, we linearly interpolated the resonant diffusion coefficients in $a$ and $j$.  The nonresonant diffusion coefficients also depend on the S-star masses. More precisely, they are a linear combination of integrals of the background distribution function, with mass-dependent coefficients. We linearly interpolated each integral term in $a$ and $j$ and combined them. Our interpolation tables extend between $j=0.001$ and $j=0.999$ and between $a=7\times 10^{-4}$ pc and $0.5$ pc. These are the boundaries of our integration domain.

The outer boundaries are purely reflective. The inner semimajor axis is purely absorptive.

At the inner $j$ boundary, stars can be disrupted
if their pericenter falls below the stellar tidal radius \citep{rees1988}, 
\begin{equation}
r_t = \left(\frac{M_{\rm BH}}{m_*}\right)^{1/3} r_*
\label{eq:rts}
.\end{equation} 
Here, $m_*$ and $r_*$ are the stellar mass and radius, respectively. These were evolved as is described in \S~\ref{sec:starEvolve}, so that the inner boundary depends on stellar properties.\footnote{In our base simulations, stars are required to spend at least one orbit within the loss cone before being removed. We also ran simulations in which stars were removed immediately once the pericenter dropped below $r_t$, and found the results to be unaffected. This was expected, since remnant stars are in the empty loss cone regime.} To avoid stars overshooting and crossing into negative angular momenta, we implemented a second, reflective inner boundary at $j=10^{-3}$.

The true tidal radius may differ by a factor of order unity from equation~\eqref{eq:rts}, based on the results of hydrodynamic simulations \citep{ryu+2020}. We used equation~\eqref{eq:rt} in any case, as there is currently no fitting formula for the tidal radius of evolved stars that is calibrated to hydrodynamic simulations.

\subsection{Stellar evolution}
\label{sec:starEvolve}
As time progresses, the stars evolve. We used both \textsc{PARSEC} isochrones \citep{bressan+2012, chen+2014, chen+2015, tang+2014, marigo+2017, pastorelli+2019, pastorelli+2020} and MIST \citep{dotter2016, choi+2016, paxton+2011,paxton+2013,paxton+2015} stellar evolution tracks\footnote{We used MIST Version 1.2 tracks with $\Omega/\Omega_{\rm crit}=0.4$.} to evolve the stellar masses, radii, and types. We find similar results with both approaches. For conciseness, we only present the results from MIST.

We present results for solar metallicity, though we find similar results for [Fe/H]=0.5. There are no existing metallicity constraints for the young stars in the Galactic center.

\subsection{Summary of model}
We now summarize our forward model for the formation of the S-star cluster,
starting from binary disruptions. The key steps are (see Fig.~\ref{fig:flow}):

\begin{figure}[h]
\centering
\includegraphics[width=\hsize]{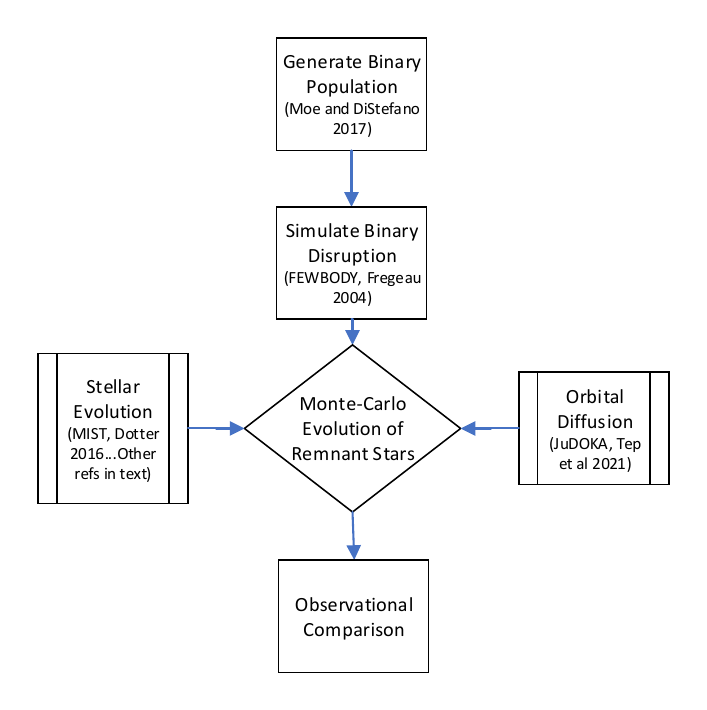}
\caption{\label{fig:flow} Key steps in our model for the production of S stars, including references and software used.} 
\end{figure}

\begin{enumerate}
    \item Generate model binaries following the assumptions in \S~\ref{sec:bprop}.
    \begin{enumerate}
        \item We only consider binaries with semimajor axes up to 15 au, as 
        wider binaries are unlikely to deposit stars into the S-star region (see equation~\ref{eq:as}).
        \item The primary mass is between $1$ and $100 M_{\odot}$. Lower-mass binaries would not produce the presently observed S stars.
    \end{enumerate}
    \item Generate close encounters with the SMBH following a Monte Carlo procedure:
    \begin{enumerate}
        \item The look-back time for the encounter is sampled uniformly between 
        $10^9$ yr ago and the present day. 
        \item The pericenter of the encounter is sampled uniformly between 16 gravitational radii ($\sim$ the tidal radius of a single star) and three times the maximum tidal radius of the binary population (equation~\ref{eq:rt}). Thus, we implicitly assume binaries are in the full loss cone regime, such that the pericenter of the center-of-mass orbit is uniformly distributed (see \citealp{merritt2013} for a discussion of loss cone physics).
        \item If the pericenter of the encounter exceeds three times the binary tidal radius (equation~\ref{eq:rt}), the binary is unlikely to disrupt, and we proceed to the next one \citep{sari+2010,generozov&madigan2020}. 
    \end{enumerate}    
    \item We simulate tidal encounters between each binary and the SMBH using the \textsc{Fewbody} code \citep{fregeau+2004}. In these encounters, the binary is initialized at 50 tidal radii on a nearly parabolic orbit with respect to the SMBH. The binary is generally split into low angular momentum bound and unbound stars (see \S~\ref{sec:bprop}).
    \begin{enumerate}
        \item We discard cases where the binary stars collide. This happens only $\sim1\%$ of the time.
    \end{enumerate}
    \item We collect all bound remnant stars with semimajor axes $< 0.05$ pc, and follow their angular momentum and energy relaxation in a fixed background, as is described in \S~\ref{sec:rlx}.  The effect of the remnant stars on the background density profile is neglected \citep{fragione&sari2018}.
    \item Stars are allowed to evolve and die (see \S~\ref{sec:starEvolve}). Furthermore, stars may be tidally disrupted by the central SMBH (see equation~\ref{eq:rts})
    \item We collect the properties of all surviving stars with semimajor axes $< 0.05$ pc and K-band magnitudes $< 18$, and compare them to observations.
    \item For each model star, we compute the probability to measure its orbital elements according to equation 1 in \citet{burkert+2023}. To account for observational bias, we exclude stars according to this probability.
\end{enumerate}

\section{Results}
\label{sec:res}
\begin{figure}[h!]
\centering
\includegraphics[width=0.85\columnwidth]{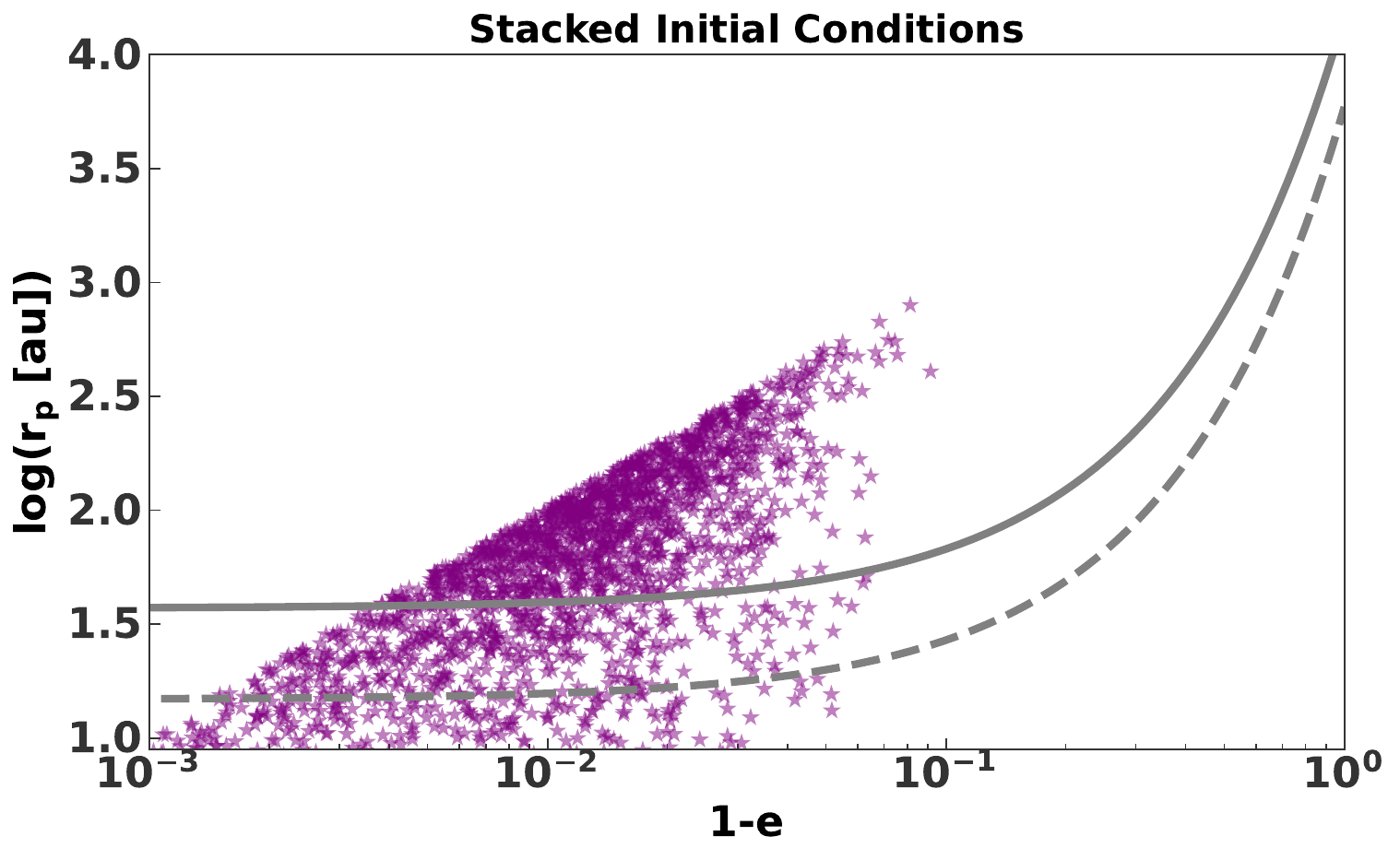}
\includegraphics[width=0.85\columnwidth]{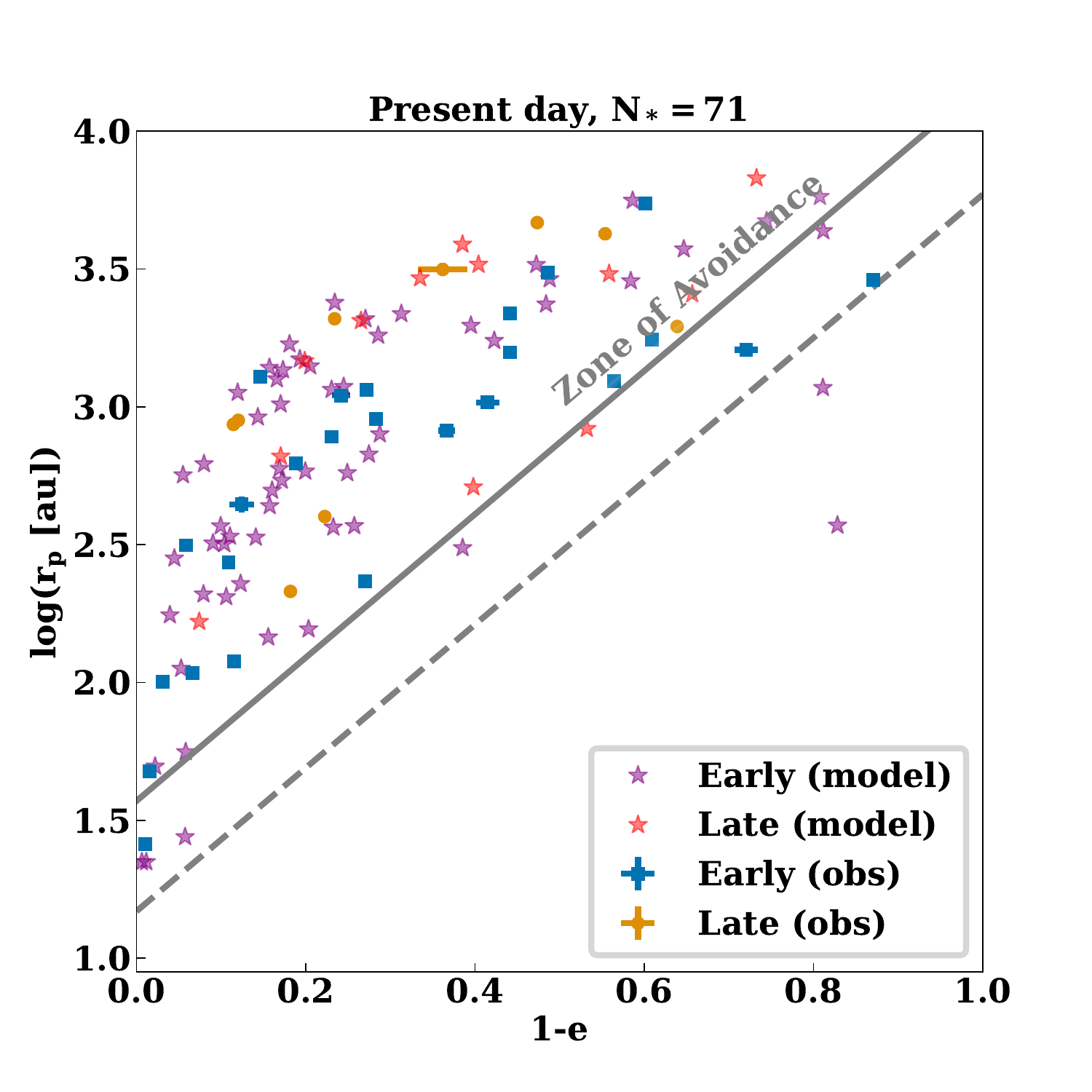}
\caption{\label{fig:res} Initial (top) and final (bottom) pericenters and eccentricities for a subsample of model stars. (Note the different x scales in the two panels.) 
The top panel shows all stars, independently of when they are deposited in the Galactic center.
The bottom panel shows the 71 model stars surviving to the present day with $K\leq 18$ and $a\leq 0.05$ pc. Early (late) stars are shown in purple (red). Model stars initially have high eccentricities and evolve toward lower ones. For comparison, we show the observed S stars in the bottom panel using blue squares for early-type stars and orange circles for late-type stars. The upper boundary of the zone of avoidance from \citep{burkert+2023} is shown as a solid gray line. The dashed gray line shows an adjusted zone of avoidance so that all S stars are outside it.}
\end{figure}

The initial conditions of bound stars with semimajor axes of $a_s \leq 0.05$ pc are shown in the top panel of Fig.~\ref{fig:res}.
As was expected, all stars start on highly eccentric orbits. The semimajor axis distribution is close to uniform, due to the approximately log-uniform semimajor axis distribution of the progenitor binaries, and our assumption of a full loss cone (see also \citealp{perets.gualandris2010}). The minimum semimajor axis is $\sim5\times 10^{-4}$ pc,  corresponding to remnants of near-contact binaries (see equation~\ref{eq:as}). After $\sim 200$ Myr, a quasi-steady state is reached in the S-star region. The final pericenters and eccentricities of surviving stars are shown in the bottom panel of Fig.~\ref{fig:res}. Purple and red stars correspond to early- and late-type stars, respectively.\footnote{In the model, the late S stars are those with $T_{\rm eff} \leq 5000$ K.} We also show the observed orbits of early and late-type S stars with blue squares and orange circles, respectively. We group observed stars of unknown type with early stars. If they were late-type stars, they would likely have been identified as such, considering CO absorption lines are prominent, and therefore easily observed with current instruments. Unless otherwise specified, here and in other observational comparisons, we exclude (i) disk S stars\footnote{These are stars with angular momentum that is comparable to the clockwise disk (see Fig.~12 in \citealp{gillessen+2017}).}, (ii) unbound S stars, and (iii) S stars with semimajor axes greater than 1.25''. The disk and the rest of the S stars differ in their eccentricity distribution, suggesting these are distinct populations that formed separately. The presence of O and WR stars in the disk also supports this. We exclude stars at large semimajor axes, as they will be strongly affected by completeness effects, complicating any comparisons to our model. Furthermore, our assumption of a full loss cone will break down at large radii (see Fig.~1 in \citealp{perets.gualandris2010}), further complicating the comparison.

Overall, we find a reasonable agreement with the zone of avoidance. Approximately $\sim14\%$ of model stars are in the zone of avoidance, as it is defined in \citet{burkert+2023}.  Similarly, in the latest orbital data, 3 of the $\sim40$ observed stars are in the zone of avoidance (S14, S18, and S23). If we adjust the zone of avoidance so that all observed stars are outside (dashed gray line in Fig.~\ref{fig:res}), then only $\sim3.9\%$ of model stars are inside the modified zone. For 40 stars, this would correspond to a $\sim20\%$ probability of no stars being in the zone of avoidance.

\begin{figure*}[h!]
\centering
\includegraphics[width=\fwidth]{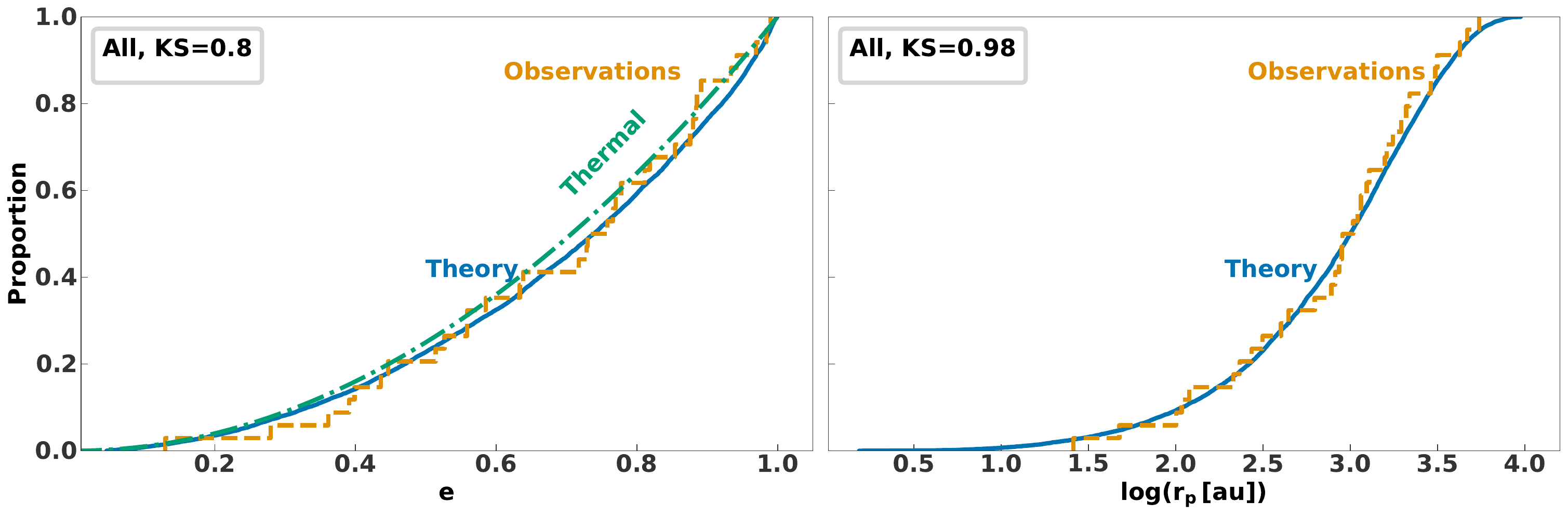}
\caption{\label{fig:1dDist} Left panel: Cumulative eccentricity distribution from the model (solid blue line), compared to the observed  S-star eccentricity distribution (dashed orange line). For reference, we also show a thermal eccentricity distribution (dash-dotted green line). Right panel: Cumulative pericenter distribution from our model and observations. The numbers in the legend are the p values from a two-sample KS test with the model and observed distributions.}
\end{figure*}

Fig.~\ref{fig:1dDist} shows the one-dimensional pericenter and eccentricity of our model stars, compared to the observed S stars. Both distributions are consistent with the observations.

\FloatBarrier

\begin{figure*}[htbp!]
\centering
\includegraphics[width=\fwidth]{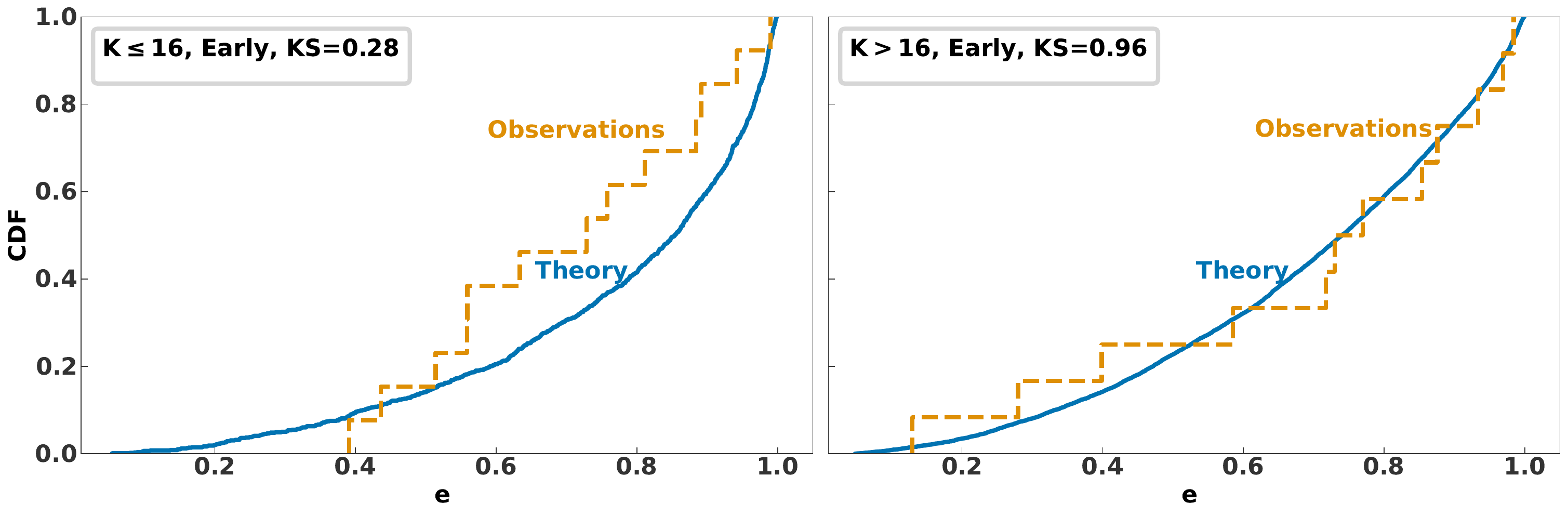}
\includegraphics[width=\fwidth]{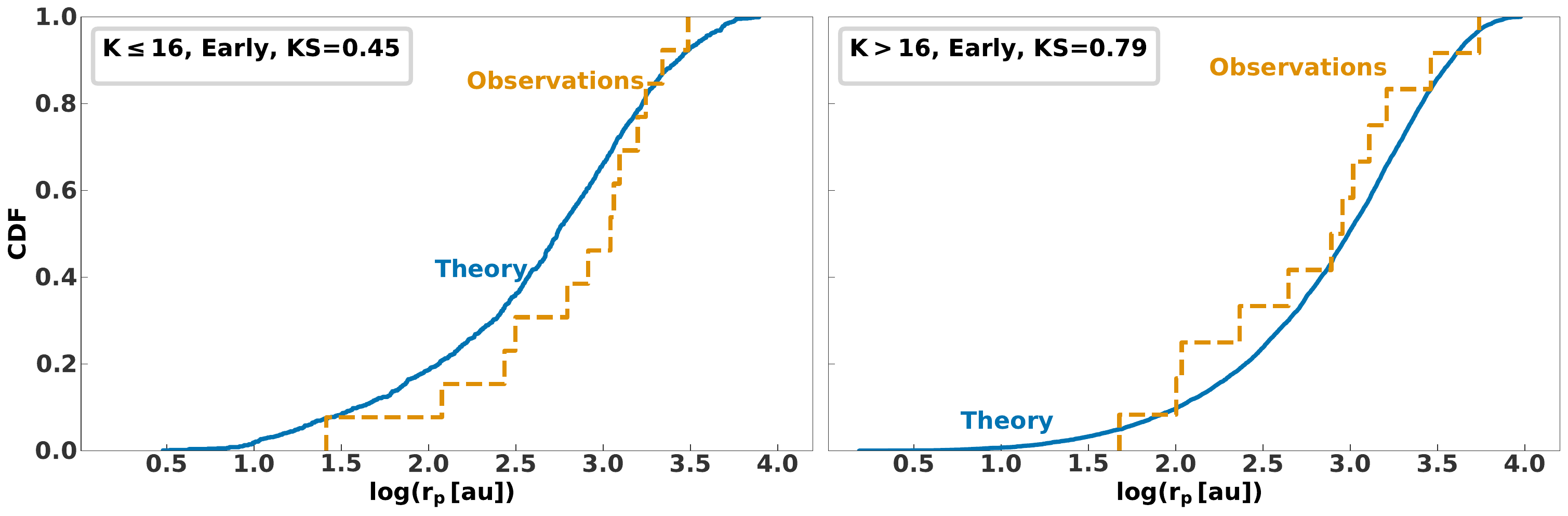}
\includegraphics[width=\fwidth]{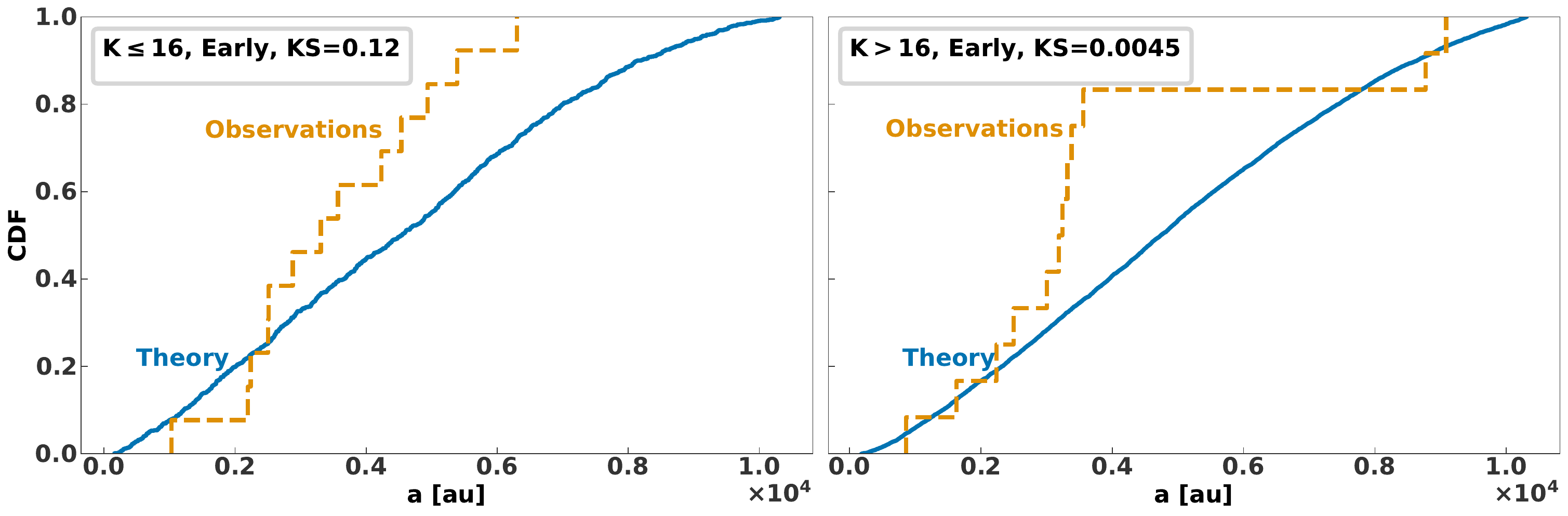}
\caption{Cumulative eccentricity, pericenter, and semimajor axis distributions for bright ($K \leq 16$, left panels) and faint ($K > 16$, right panels) early S stars.}
\label{fig:brightFaint} 
\end{figure*}

Fig.~\ref{fig:brightFaint} shows the orbital properties of early S stars with $K \leq 16$ and $K>16$.  This magnitude roughly divides the observed sample in half. The model eccentricity and pericenter distributions of each group are consistent with observations, though the bright stars have a somewhat superthermal eccentricity distribution. 
The faint stars' semimajor axis distribution is too skewed toward large 
semimajor axes, compared to observations. In part, this is due to the assumed model for the progenitor binaries: as the primary mass decreases, the period distribution gradually shifts from approximately log-uniform to log-normal (with a peak near $10^5$ days). We also note that faint stars at large semimajor axes are the most difficult to detect, and this comparison is subject to observational bias.

\FloatBarrier

We now compare the stellar properties of model stars with observations. In the model, $\sim20\%$ of stars are late-type. This is consistent (within Poisson uncertainties) with the observed fraction of $\sim 26\%$. However, we note that we have not included disruptions of old ($\gsim 10^{10}$) stars that make up the bulk of the nuclear stellar disk, and that can increase the population of evolved stars in the S-star region.

\begin{figure}
\centering
\includegraphics[width=\fwidth, height=\fheight]{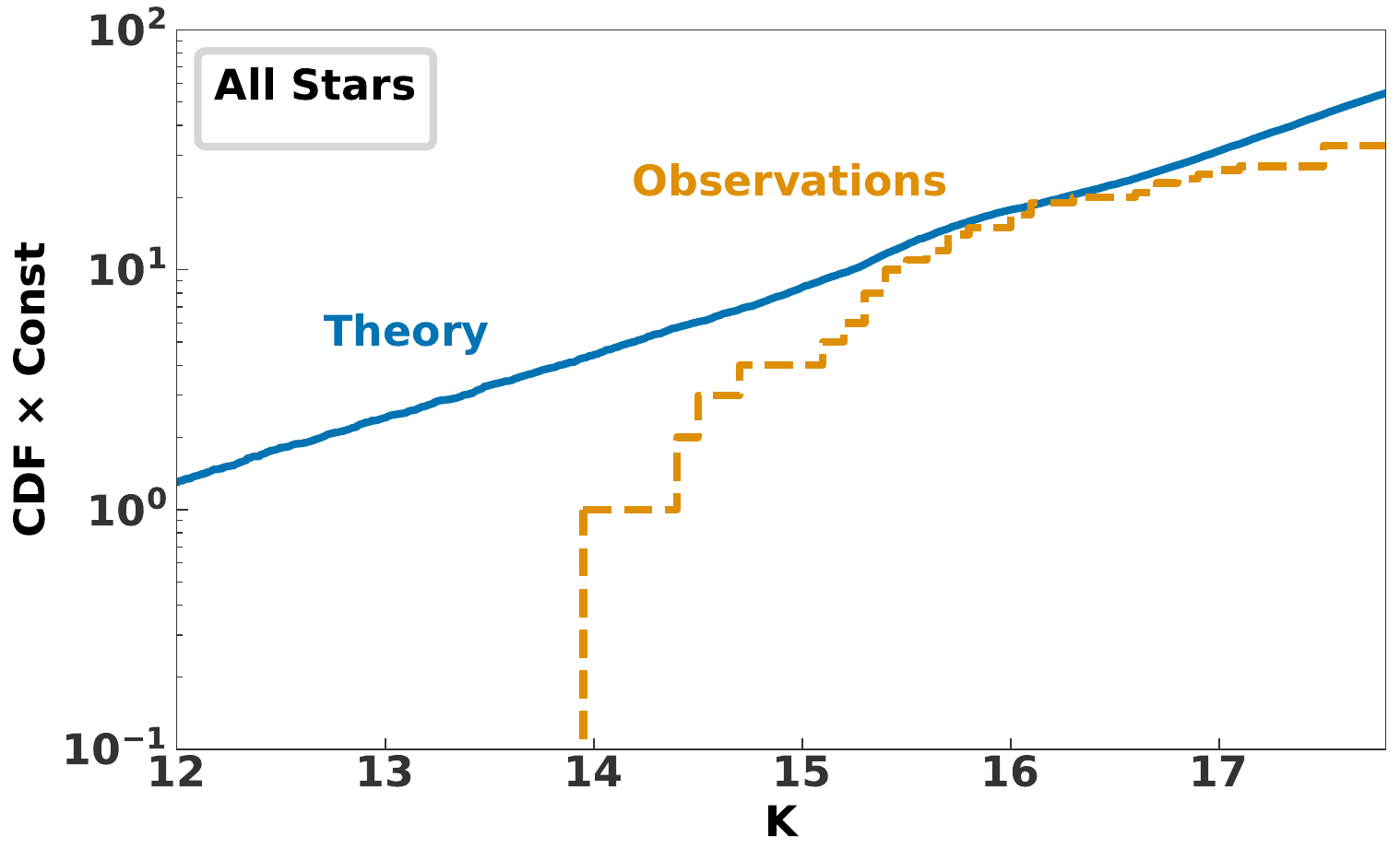}
\includegraphics[width=\fwidth, height=\fheight]{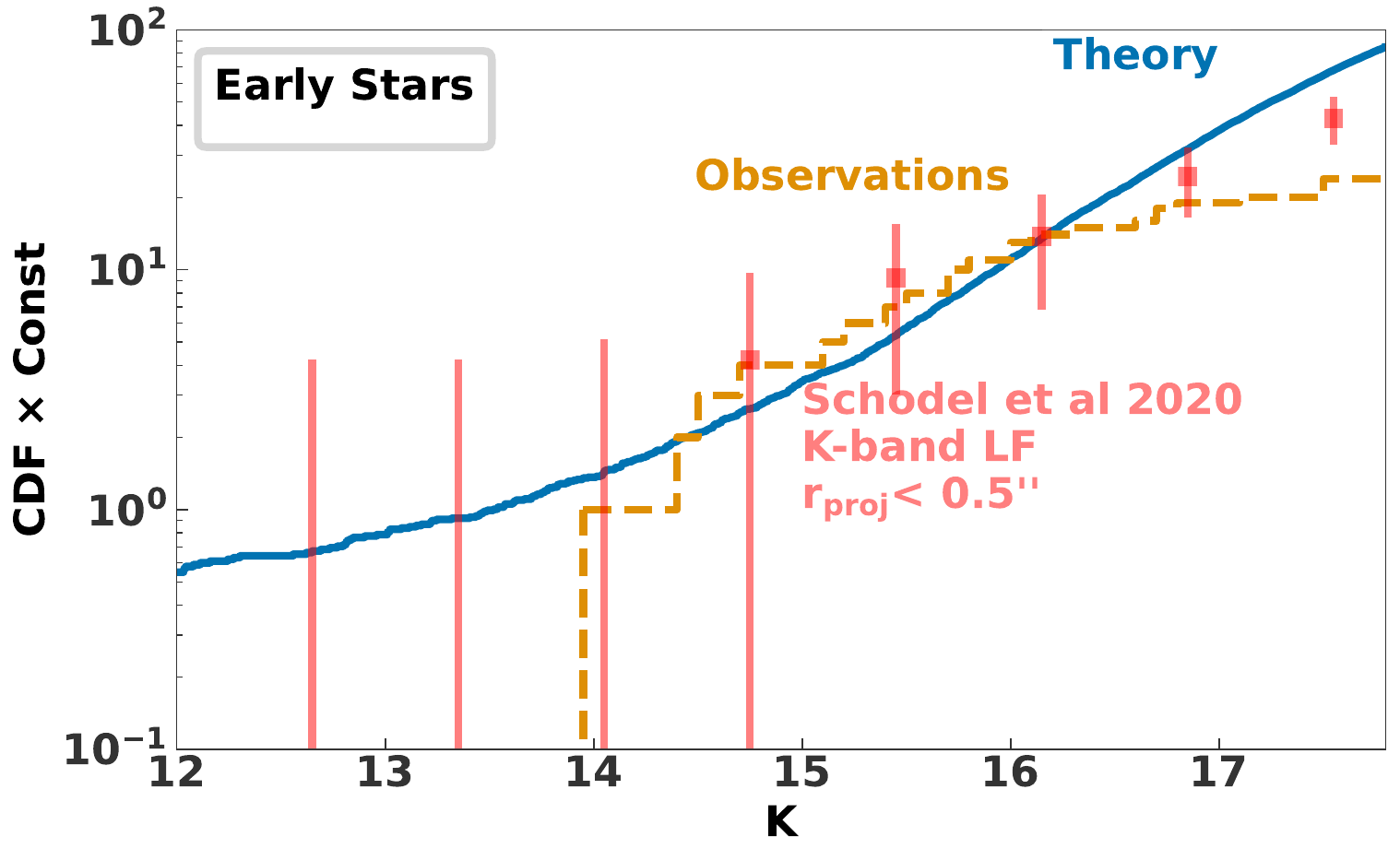}
\caption{\label{fig:kmag}  Top panel: KLF of all model (blue) and observed (dashed orange) S stars. Bottom panel: KLFs for model (blue) and observed (dashed orange) early S stars (with orbits). The red line is the KLF of early stars in the central 0.5'' from \citet{schodel+2020} (see Appendix~\ref{sec:sKLF} for details). The luminosity functions are normalized to match at a K-band magnitude of 16.2.}
\end{figure}

We also compare the K-band magnitudes of our model stars to observations (Fig.~\ref{fig:kmag}). In practice, we used the JWST F210M magnitude from MIST isochrones \citep{dotter2016, choi+2016, paxton+2011,paxton+2013,paxton+2015} to approximate the K-band magnitude of our model stars. We also assumed an extinction of 2.42 \citep{fritz+2011} and a Galactocentric distance of 8.3 kpc \citep{gravity+2021_dist}. We neglected corrections from differential extinction ($\sim 0.3$ mag; \citealp{schodel+2010extinction,do+2013}).

The top panel of Fig.~\ref{fig:kmag} compares the overall KLFs from theory and observations. We rescaled the luminosity functions to match at $K=16.2$. There are significant discrepancies between the model and observed S-star KLF at both the faint and bright ends. The former can plausibly be explained by completeness effects. The latter is due to $\sim 3$ bright giants with $K < 14$ that are not observed, and is of marginal significance.
We note that collisions with black holes cannot remove these bright giants (see \S~\ref{sec:coll}).

The bottom panel of Fig.~\ref{fig:kmag} compares the model and observed KLF of early-type stars (blue and orange). Here, there are no large differences at the bright end, while the model has many more stars with $K>16$. This discrepancy may again be due to completeness effects. 
Unfortunately, this is difficult to quantify, as the observational selection function is highly nontrivial (the orange KLF only includes stars with measured orbits). The model more closely matches overall KLF from \citet{schodel+2020} (light red points; see Appendix~\ref{sec:sKLF} for details), which should be more complete, as even stars without orbits are included.

However, the progenitor binaries are more likely to follow a present-day mass function than an initial mass function (as assumed here). A present-day mass function would steepen the model KLF, and exacerbate the tension with observations (see \S~\ref{sec:add}).

\FloatBarrier

\begin{figure}
    \centering
    \includegraphics[width=\fwidth, height=\fheight]{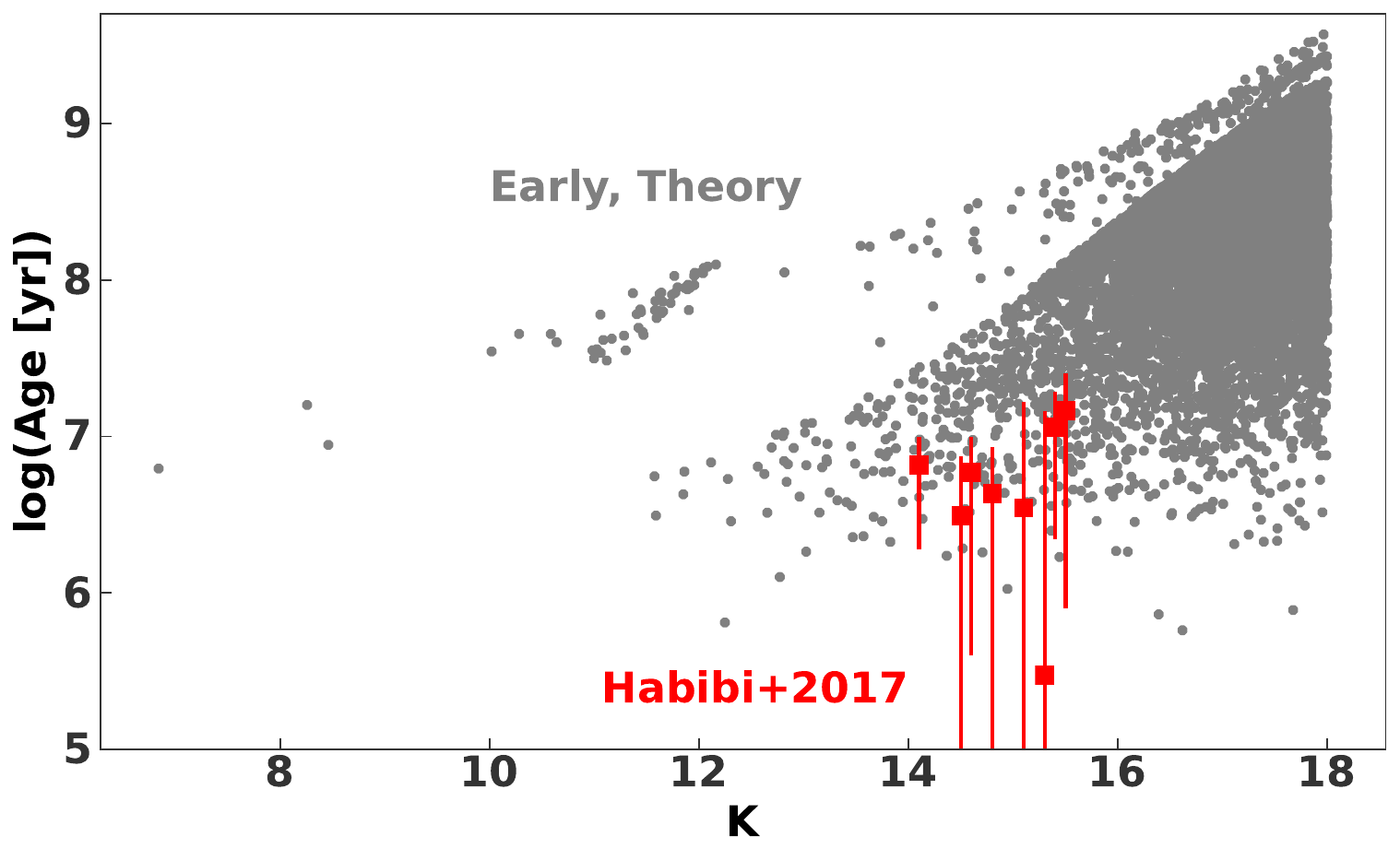}
    \includegraphics[width=\fwidth, height=\fheight]{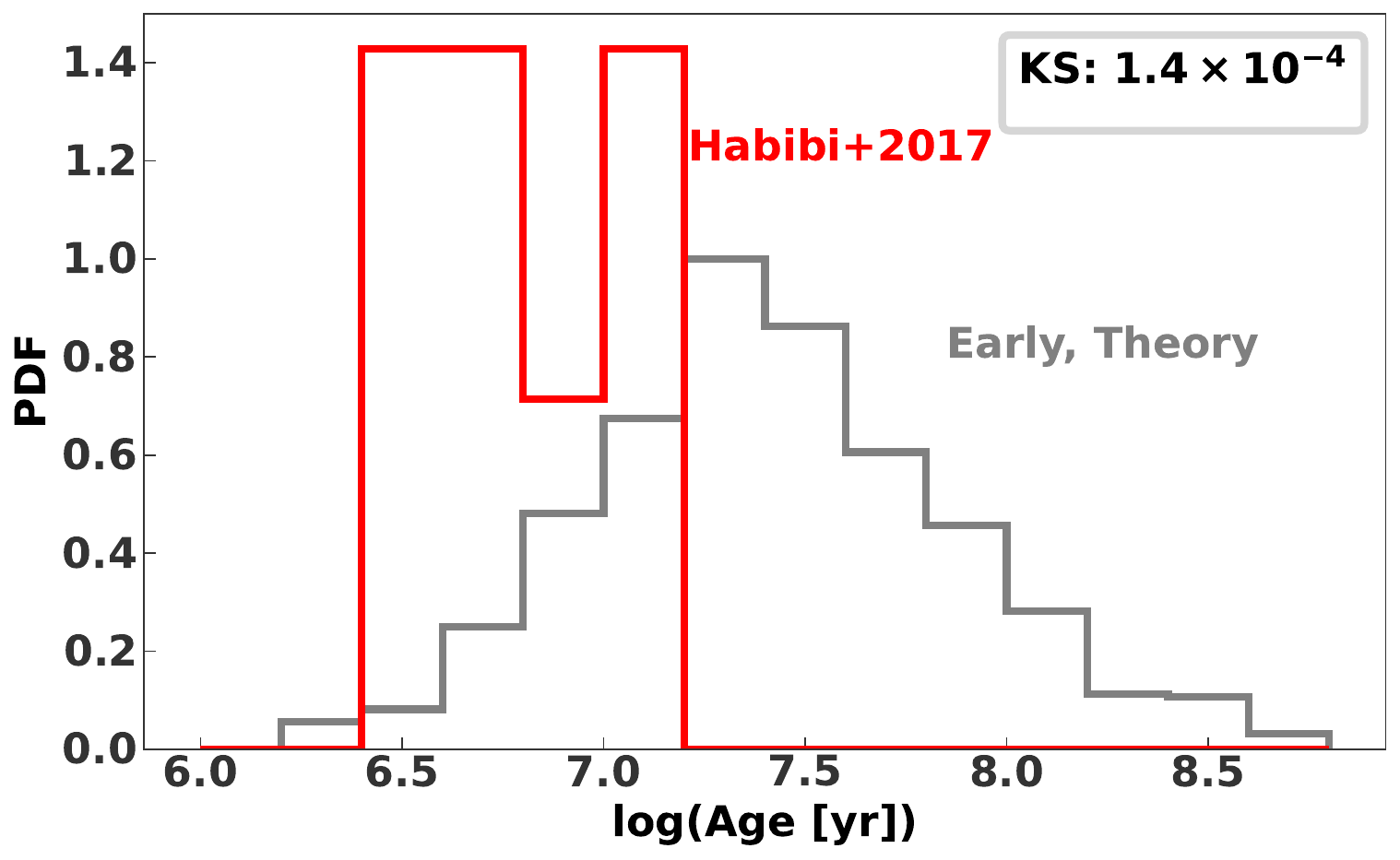}
    \caption{\label{fig:habibi} Top panel: Age versus K-band magnitude for eight bright S stars (red; \citealp{habibi+2017}), compared to early model stars (gray). Bottom: Age distribution of  \citet{habibi+2017} stars and model stars with comparable magnitudes (see text for details).}
\end{figure}

The top panel of Fig.~\ref{fig:habibi} shows the K-band magnitude as a function of age for the eight bright ($K \leq 15.5$) S stars from \citet{habibi+2017} and for our model stars, which appear significantly older. To better compare the age distributions, we selected a subset of model stars with comparable K magnitudes. First, we selected all model stars within 0.1 mag for each of the \citet{habibi+2017} stars. Then, we sampled 100 stars (with replacement) from each of these subsets.  The gray line in the bottom panel of Fig.~\ref{fig:habibi} shows the stacked age distribution of the selected stars. On average, model stars are significantly older, even after accounting for the observational bias toward bright stars. This may suggest that the brightest S stars have a distinct origin or that the star formation rate (and delay time distribution between star formation and disruption) is nonuniform. Finally, there is a degeneracy between the age and metallicity (see Appendix~\ref{sec:metal}), such that the S stars may in fact have subsolar metallicity and be significantly older. Currently, there are no measurements of metallicity for the S stars, but future observations of the S stars with ERIS and NIRSpec will constrain this. 
A priori, this explanation is less likely, considering the nuclear stellar disk and nuclear star cluster are mostly metal-rich with small metal-poor populations \citep{do+2015, feldmeier+2017Metal,rich+2017,nandakumar+2018,schultheis+2019,schodel+2020,schultheis+2021,nieuwmunster+2024,NoguerasLara+2024}.

\begin{figure}[h]
    \centering
    \includegraphics[width=\hsize]{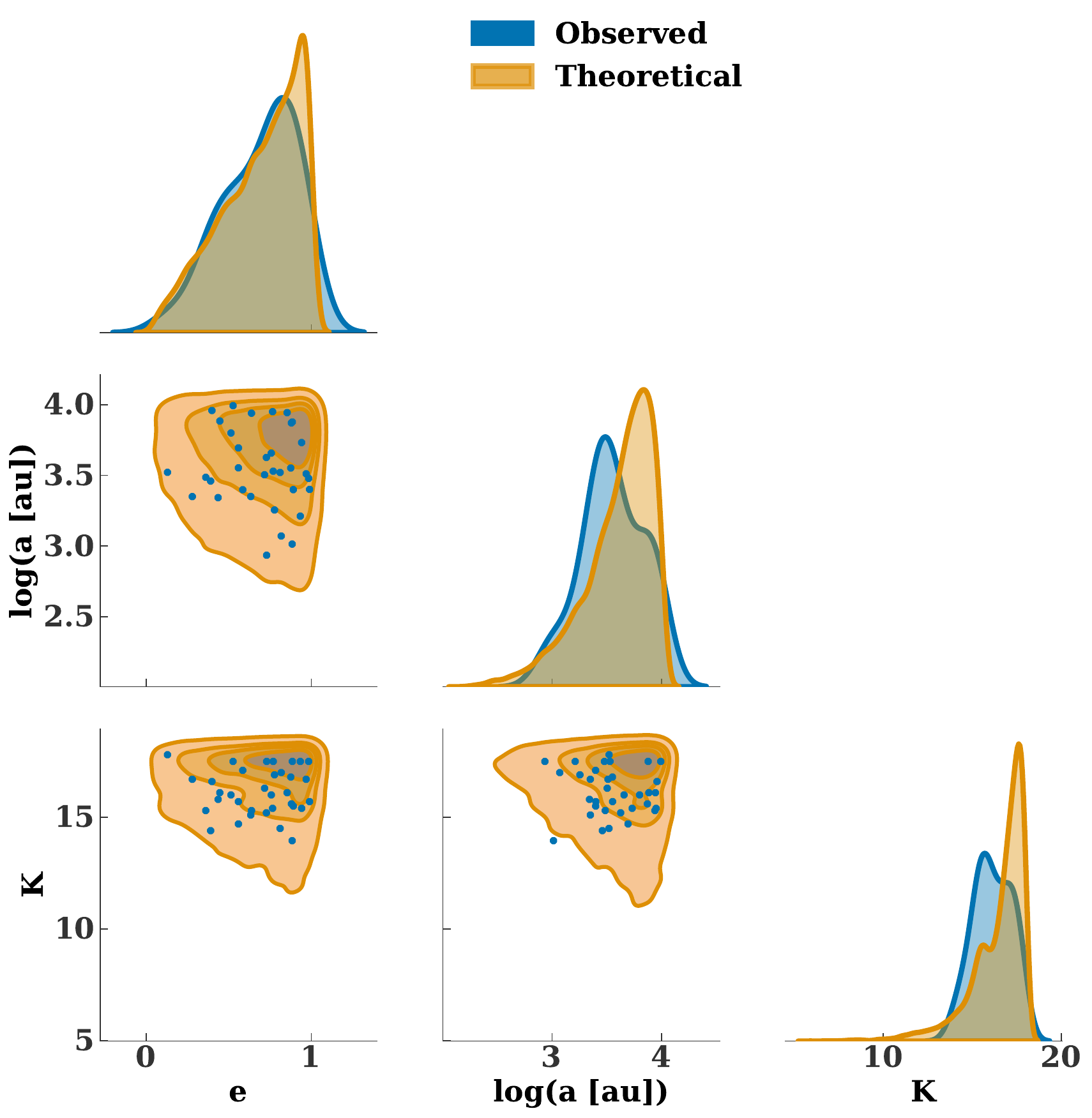}
    \includegraphics[width=\hsize]{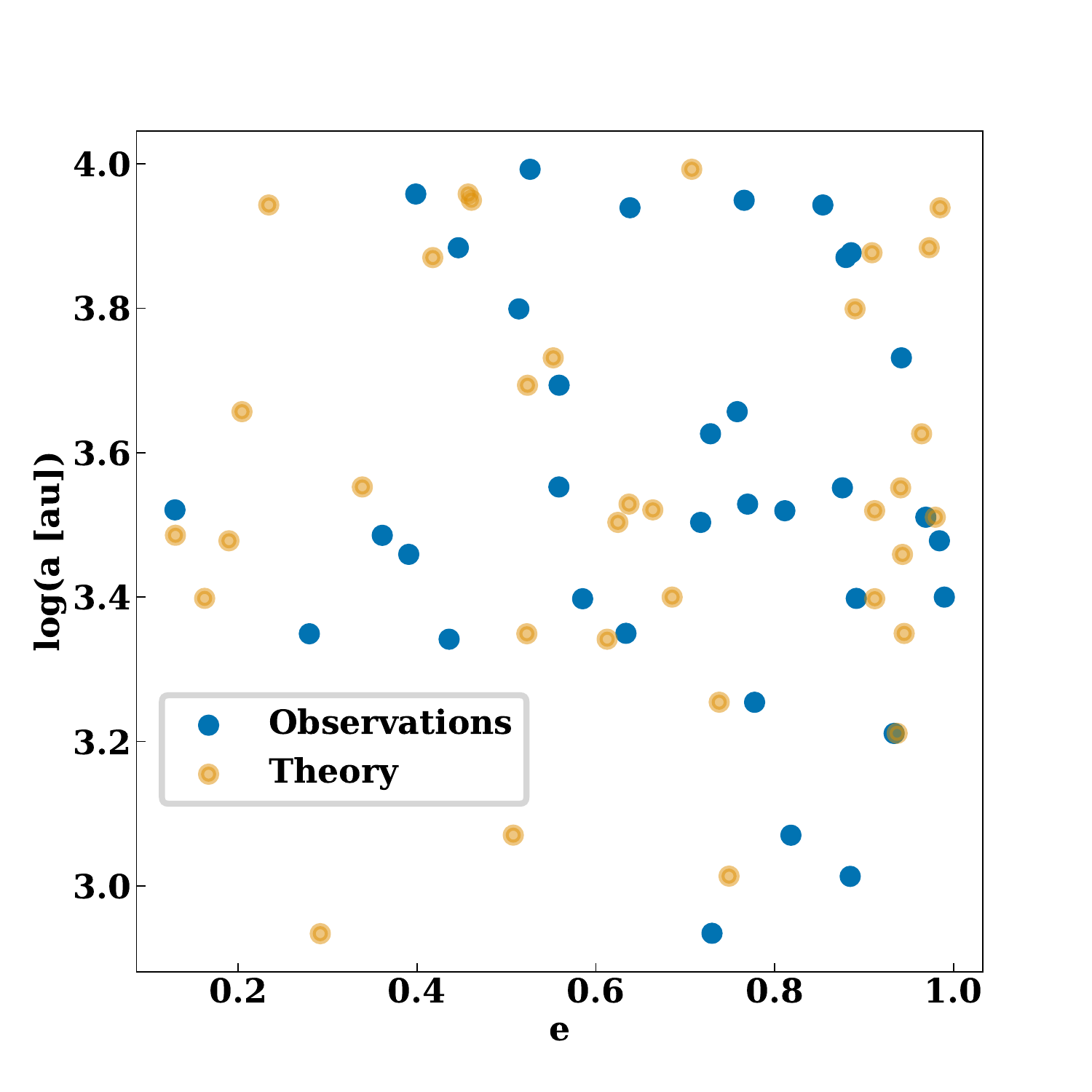}
    \caption{Top panel: Corner plot showing distributions of semimajor axis ($a$), eccentricity ($e$), and $K$ magnitude from theory (orange) and observations (blue). Bottom panel: Eccentricity versus semimajor axis for the observed S stars, and for the model stars that are closest in the semimajor axis.}
    \label{fig:corner}
\end{figure}

Finally, Fig.~\ref{fig:corner} shows a corner plot of $a$, $e$, and $K$. Once again, we see a good agreement between the observed and theoretical eccentricity distributions. At the same time, the model produces too many faint stars, and too many stars at large semimajor axes (though again we caution that faint stars at large semimajor axes are the most difficult to detect, and the latter discrepancy may be due to observational bias). In the observations, many of the lowest-eccentricity stars are in a narrow band of semimajor axes ($3.3 \lsim \log(a / {\rm au}) \lsim 3.55$). This feature is not apparent in the model. However, there are not enough stars to conclude that this is a statistically significant difference. Furthermore, the absence of this feature may simply reflect the differences in the one-dimensional semimajor axis distributions. To test this, we selected the model star with the closest semimajor axis for each observed S star.
The structure of these model stars in the $a-e$ space is more similar to the observed one, as is shown in the bottom panel of Fig.~\ref{fig:corner}.
Interestingly, there is a band of low-eccentricity stars near the Galactocentric radius where the scalar resonant relaxation time\footnote{$t_{\rm RR} = \frac{1}{\int_{j_{\rm lc}}^{1} D_{\rm jj, RR} 2 j d j}$ \citep{bar-or.fouvry2018}} is minimized ($\log(a / {\rm au})\approx 3.24$). This is where angular momentum relaxation is most efficient.

\section{Critical discussion of model assumptions}
\label{sec:add}
Here, we critically examine a few assumptions about the progenitor binary population, and how they affect observational comparisons. Firstly, the primary mass of each binary is drawn from an initial mass function. However, a steeper present-day mass function is more realistic, considering the shorter lifetimes of massive stars. 
Secondly, binaries disrupt at a constant rate and are allowed to disrupt immediately after formation. In reality, binaries may need some time to reach disruption, especially if they form on circular orbits. As is discussed below, delays and a present-day mass function exacerbate the tension with the observed KLF.

We also discuss the effect of stellar collisions and repeated, individually nondisruptive tidal encounters between the stars and the central SMBH. These effects are not included in our base model, but do not affect the results.

\subsection{Mass function and disruption times}
In our base model, the primary mass of each binary is drawn from a Kroupa initial mass function. However, the mass function of disrupting binaries will likely be more bottom-heavy, due to the longer lifetimes of low-mass stars. 

\begin{figure}
\centering
\includegraphics[width=\fwidth, height=\fheight]{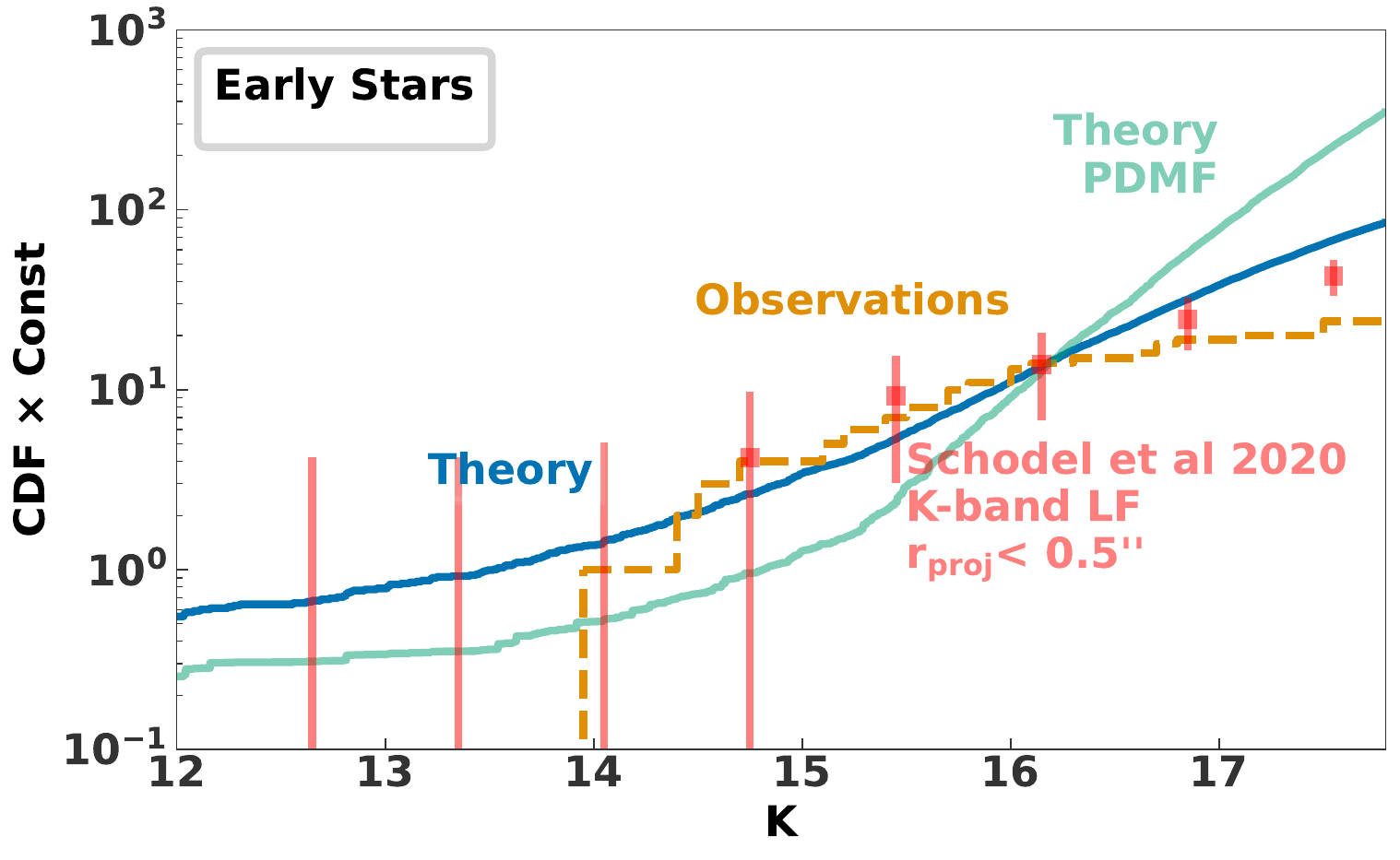}
\caption{\label{fig:pdmf1} K-band luminosity function accounting for the dependence of the binary disruption probability on stellar lifetime (light green line). Effectively, this means that the progenitor binaries are drawn from a present-day mass function rather than an initial mass function. The other lines are the same as in Fig.~\ref{fig:kmag}.}
\end{figure}

Here, we account for this effect by re-weighting probability distributions. In particular, each remnant star is weighted by the main-sequence lifetime of the primary star in the progenitor binary. This causes the KLF to steepen, in tension with observations, as is shown in Fig.~\ref{fig:pdmf1}.

This result depends on the star formation history in the Galactic center, particularly within the region of the nuclear stellar disk, where disrupting binaries likely originate (see Fig.~3 in \citealp{perets+2007}). \citet{schodel+2023} recently showed that the star formation history in the nuclear stellar disk is highly nonuniform: $\gsim70\%$ of the stars formed more than $10$ Gyr ago, $\sim15\%$ formed $\sim1$ Gyr ago, and up to $10\%$ formed in the last tens to hundreds of millions of years.
This suggests a star formation rate rising toward the present day (since a 
comparable number of stars were formed within the last $\sim10^8$ yr and $\sim10^9$ yr). 
Furthermore, the star formation rate is likely correlated with the binary disruption rate, since both are correlated with the molecular gas density. 

\begin{figure}
\centering
\includegraphics[width=\fwidth, height=\fheight]{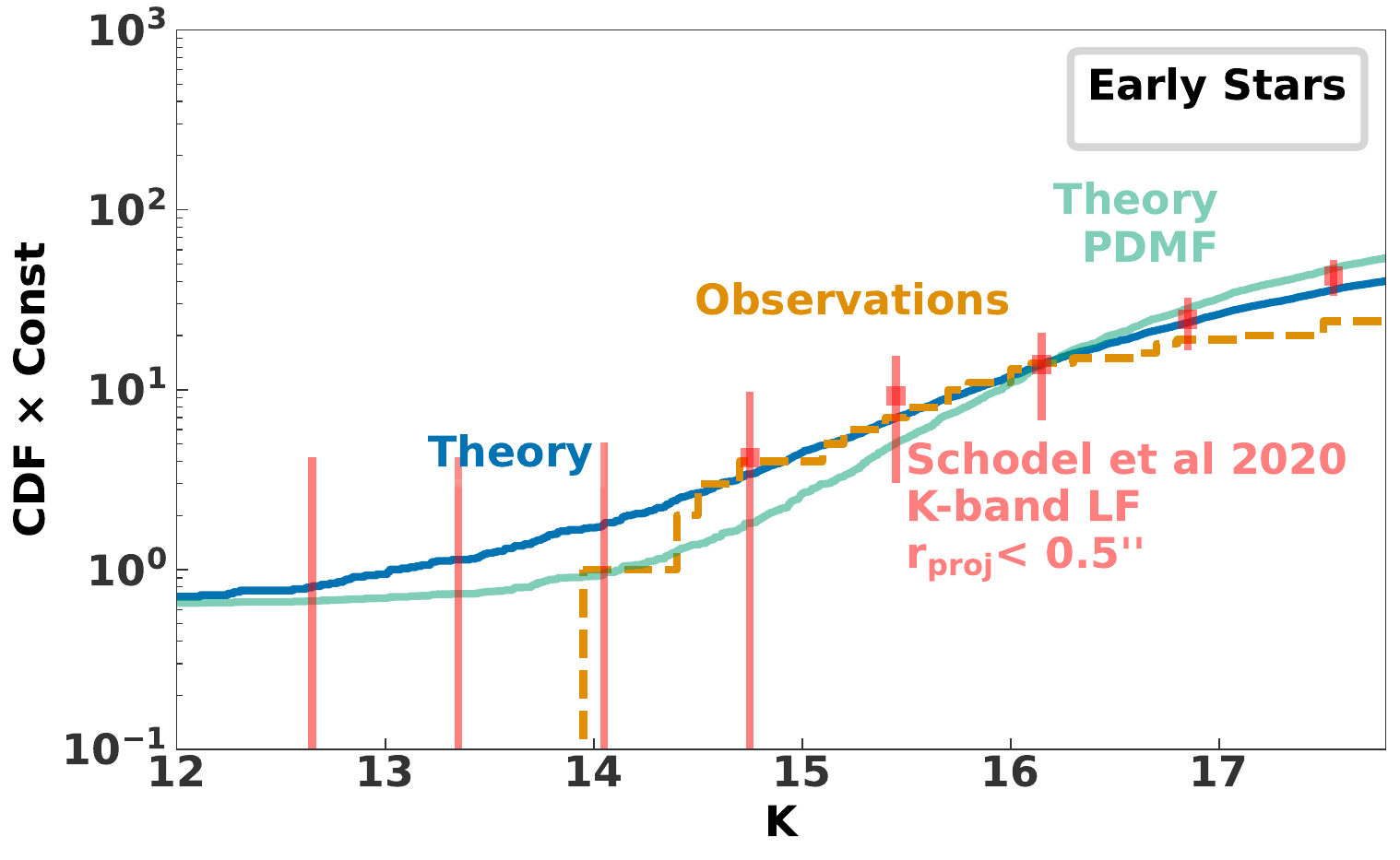}
\caption{\label{fig:pdmf2} K-band luminosity functions, after discarding stars formed more than $10^8$ yr ago. The blue (light green) line corresponds to binaries drawn from an initial (present-day) mass function.}
\end{figure}

Motivated by these observations, we discarded stars that formed more than $10^8$ yr ago to see the effect of a star formation history rising to the present day. Fig.~\ref{fig:pdmf2} shows the KLF after this cut, with and without re-weighting for stellar lifetimes.
In the former case, massive stars receive a weight equal to  
their main-sequence lifetime divided by $10^8$ yr. Stars with main-sequence lifetimes longer than $10^8$ yr receive a weight of unity. This ameliorates, but does not entirely remove, the tension with observations, as is shown in Fig.~\ref{fig:pdmf2}. Furthermore, the model eccentricity distribution is in tension with observations, though this can be resolved if the background black holes are each 20 $M_{\odot}$ instead of 10 $M_{\odot}$. 

\begin{figure}
\centering
\includegraphics[width=\fwidth, height=\fheight]{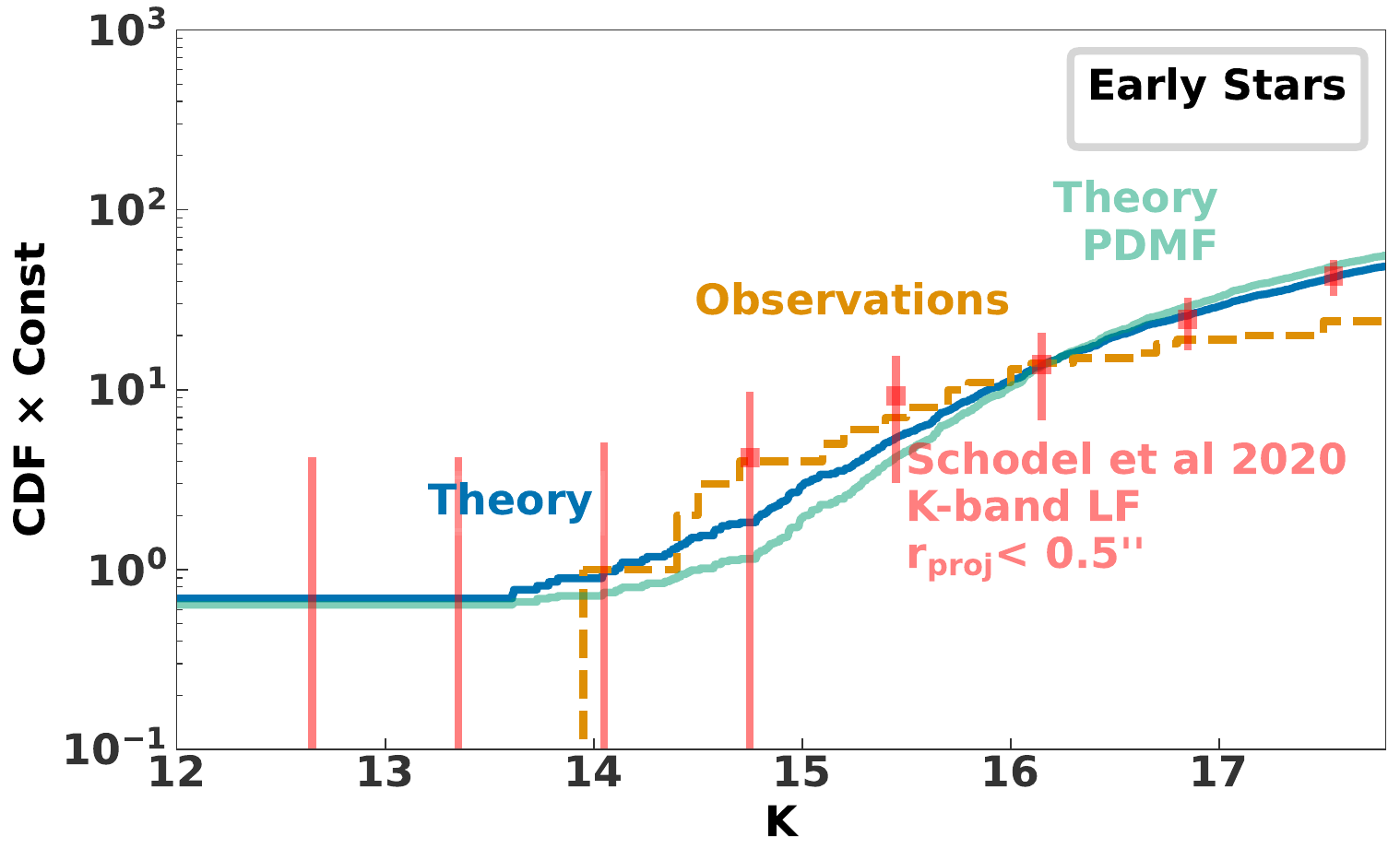}
\caption{\label{fig:ktmin} Same as Fig.~\ref{fig:pdmf2}, except we discarded 
stars that were less that $10^7$ yr old at the moment of binary disruption.}
\end{figure}

The results also depend on the delay time distribution between star formation and disruptions. In the base model, binaries can be disrupted immediately after formation. More realistically, binaries will take some time to reach the loss cone, especially if they form on circular orbits. Imposing a minimum age at disruption tends to steepen the KLF, as is illustrated in Fig.~\ref{fig:ktmin}. This shows the KLF after (i) discarding stars older $10^8$ yr and (ii) discarding stars injected into the Galactic center less than $10^7$ yr after formation. 

The better agreement between theory and observation in Fig.~\ref{fig:pdmf2} suggests that the early S stars are primarily sourced by binaries formed within the last $\sim 100$ Myr. Nonetheless, a slight tension remains, especially for realistic delay times between star formation and disruption, as in Fig.~\ref{fig:ktmin}. In this case, there is a $\sim$2$\sigma$ tension between the observed (orange) and theoretical (turquoise) KLF  between $K=14$ and $K=16$.\footnote{We have not included fainter stars in this comparison, as the observations may be incomplete. In fact, we find that at faint magnitudes the theoretical KLF agrees better with the KLF from \citet{schodel+2020}, which is less sensitive to completeness effects (see the discussion in \S~\ref{sec:res}).} Potentially, this tension could be resolved via a bursty star formation rate that is more skewed toward the present day.
This may also help with the tension between model and observed ages among the bright S stars (see Fig.~\ref{fig:habibi}).

Alternatively, the tension with the observed KLF may be resolved with a top-heavy initial mass function. The mass function required itself depends on the binary injection history. In the last case (Fig.~\ref{fig:ktmin}), an initial mass function of $dN / dm \propto m^{-0.2}$ reduces the tension to 1$\sigma$.

\subsection{Collisions}
\label{sec:coll}
In principle, stars may suffer mass loss or destruction via collisions with other stars or compact objects. Alternatively, star-star collisions can lead to mergers and the buildup of massive stars \citep{spitzer&saslaw1966,duncan&shapiro1983,murphy+1991,genzel+1996,davies+1998,portegies-zwart+1999,gurkan+2004,freitag+2006Coll,dale+2009,rose+2023}.

We estimated the impact of destructive black-hole star collisions. In particular, we computed the orbit-averaged collision rate between the remnant stars and black holes in our Monte Carlo models:

\begin{equation}
\Gamma_{\rm coll}(r) = n_{\rm bh}(r) v(r) \pi r_*^2 \left(1 + 2 \frac{G (m_{\rm bh} + m_*)}{v(r)^2 r_*}\right),
\label{eq:collRate}
\end{equation}
where $m_*$ and $m_{\rm bh}$ are the stellar and BH masses, respectively, 
$r_*$ is the stellar radius, $v(r) \approx \sqrt{G M_{\rm bh} /r}$ is the local velocity dispersion, and $n_{\rm bh}(r)$ is the BH density (see Eq.~\ref{eq:background}). The orbit average was calculated from $10^4$ points along the orbit, equally spaced in eccentricity anomaly ($EA$). Each point is weighted by $\left(1 - e \cos(EA)\right)$ in the average.

For each Monte Carlo timestep, $\delta t$, we generated a random number between 0 and 1, and recorded a collision if it was less than the expected number of collisions, $<\Gamma_{\rm coll}> \delta t$.

Removing stars that experience collisions has a minor impact on the results. 
Most notably, the fraction of late-type S stars decreases from $\sim20\%$ to 
$\sim16\%$.  In fact, the impact of collisions may be overestimated, considering that deeply penetrating encounters are necessary to significantly dim giant stars \citep{dale+2009}. 

\citet{rose+2023} have shown that star-star collisions in the S-star region can lead to the buildup of massive stars (but this would require nondisruptive collisions).
We leave this effect to future studies, though it may have a major impact on the comparisons to the observed KLF and ages. Additionally, stellar collisions can lead to transients that look like tidal disruption events or supernovae \citep{balberg+2013,amaro-seoane2023,ryu+2024a,dessart+2024}.

\subsection{Repeated tidal encounters with the SMBH}
So far, we have only considered the removal of stars due to full tidal disruptions by the central SMBH. However, stars can also be removed due to multiple partial tidal disruptions. Furthermore, stars can become swollen and disrupted due to tidal heating \citep{alexander&morris2003}. To experience such effects, the star needs to pass within a factor of a few of the central SMBH \citep{guillochon&ramirezruiz2013,li&loeb2013,mainetti+2017,ryu+2020part,wenbinlu+2020,generozov2021}. To test how these effects would change the results, we multiplied the tidal radius of each star by two. We found that this has no significant effect on the comparisons previously discussed.

\section{Fixed age model (eccentric disk)}
\label{sec:imp}
So far, we have considered models in which stellar ages and disruption times  can vary freely. We now discuss models in which all stars are injected at similar times in the recent past. Such an impulsive burst of disruptions may be triggered by a secular gravitational instability in the $\sim2.5-8$ Myr old \citep{levin&beloborodov2003, paumard+2006, bartko+2009, lu+2013, yelda+2014} clockwise disk if it starts in a lopsided, eccentric configuration \citep{madigan+2009, generozov&madigan2020,rantala&naab2023}. This configuration can arise from the tidal disruption of a molecular cloud \citep{bonnell&rice2008,gualandris+2012,gnm+2022}, or via an SMBH merger in the Galactic center \citep{akiba+2024}.\footnote{Mergers of clusters and IMBHs and/or SMBHs with the Galactic center may also form or shape the S-star cluster via other processes (e.g., scattering, dynamical friction) \citep{kimfiger+2004,fragione+2017,cao+2024}.} Interactions between the misaligned young disks can also trigger a burst of binary disruptions \citep{lockmann+2008}.

Previously, \citet{generozov&madigan2020} and \citet{tep+2021} analyzed the relaxation of the S-stars in such scenarios. 
The former found that $10 M_{\odot}$ BHs would be able to reproduce the observed S-star eccentricity
distribution, by modeling angular momentum relaxation at a fixed semimajor axis of 0.01 pc. However, the resonant relaxation time will be significantly longer in the outer part of the S-star cluster. \citet{tep+2021} accounted for the S stars' semimajor axis distribution and found that reproducing the observed S-star eccentricity distribution 
requires black holes above $100 M_{\odot}$ at $2\sigma$ confidence, though this constraint depends on the S-star sample (i.e., this is the constraint including late-type and disk S stars).

We performed a similar analysis, assuming that stars are injected into the Galactic center all at once. First, for consistency with \citet{tep+2021}, we turned off (i) the loss cone, (ii) stellar evolution, and (iii) semimajor axis diffusion. Effects (i) and (ii) will be negligible for the S stars over million-year timescales. However, we show that the loss cone can affect black hole mass constraints by a factor of $\sim2$.

We also matched the initial conditions in \citet{tep+2021}, assuming that all stars were injected into the Galactic center 7.1 Myr ago, the mean age of eight bright S stars from \citet{habibi+2017}.\footnote{\citet{tep+2021} used the age measurements in \citet{habibi+2017} for the eight bright S stars, and 7.1 Myr otherwise.} The initial angular momentum distribution is a Gaussian with a mean of 0.2 ($e=0.98$) and a standard deviation of 0.02.

We present results for both the density profiles in \citet{tep+2021}:
\begin{align}
&\rho_{\rm bh} = 3.6 \times 10^6  \left(\frac{r}{0.1 \,{\rm pc}}\right)^{-1.8} M_{\odot}\, {\rm pc}^{-3} \nonumber\\
&\rho_{\rm *} =  9.4 \times 10^5 \left(\frac{r}{0.1 \,{\rm pc}}\right)^{-1.5} M_{\odot}\, {\rm pc}^{-3},
\end{align}
and our fiducial density profile (Eq.~\ref{eq:background}), which is in better agreement with the latest constraints from GRAVITY \citep{gravity+2024extendedmass} and which has a factor of $3$ fewer BHs within 0.05 pc.

Using the same S-star sample and background profile as \citet{tep+2021}, we find good agreement with their results: black hole masses $<100 M_{\odot}$ can be ruled out. However, the full sample includes late-type stars and stars at larger scales that may be unrelated to binary disruptions. For example, eight of the S stars are members of the clockwise disk. For the
25 early, non-disk S stars with semimajor axes within 0.05 pc, 
only black holes with masses $<17 M_{\odot}$ can be ruled out. For the same subsample and our fiducial density profile, masses $<50 M_{\odot}$ can be ruled out. Finally, the lower limit is reduced to $\sim24 M_{\odot}$, with a loss cone.

Overall, explaining the entire S-star population via binaries from 
the young clockwise disk requires a background with more massive black holes. 
There is also tension between the mass functions of the disk and S stars that may require an unusual mass ratio distribution within the disk to reconcile \citep{generozov2021}. Moreover, it is not clear whether the conditions for the eccentric disk instability could arise with steep background density profiles (e.g., Eq.~\ref{eq:background}; see the discussions in \citealp{madigan+2009} and \citealp{gnm+2022}). On the other hand, efficient angular momentum relaxation of the S stars likely requires a steep black hole cusp. Detailed modeling of disk binary disruptions suggests that they can reproduce the zone of avoidance, but remnant stars will be skewed toward large semimajor axes compared to the observations.\footnote{\citet{generozov&madigan2020} reproduced the S-star semimajor axis distribution but considered only a single disruptive encounter per binary and focused on the region within 0.03 pc. They also used a different approach for generating binaries, sampling first a thermal eccentricity and then a semimajor axis from a log-uniform distribution, resulting in an overall binary semimajor distribution that is flatter than log-uniform and likely less realistic than the empirically motivated model here. Finally, we caution that alternative disk initial conditions (e.g., a larger inner edge) may better reproduce the observed S-star semimajor axis distribution.}

\section{Summary and discussion}
\label{sec:conc}
We show that many properties of the observed S-star cluster in the Galactic center, including the recently discovered zone of avoidance, can be explained by a combination of isotropic binary disruptions, and relaxation in the Galactic center. Our focus is on explaining the classic, isotropic S-star cluster. In our model, the young disks would have a different origin that we leave to future work.

Our main conclusions are summarized as follows.
\begin{enumerate}
    \item In the full-loss-cone regime, the remnant stars from massive binary disruptions will have an approximately uniform semimajor axis distribution. This naturally explains the lack of S stars in the zone of avoidance. Empty loss cone models would place too many stars in this region.
    \item The overall orbital eccentricity and pericenter distribution of the S-star cluster is well reproduced in our models, via relaxation processes. 
    We have only considered nonresonant and resonant relaxation from a background population of stars and $10 M_{\odot}$ black holes. Potentially, an AGN disk
    may also perturb the S-star orbits, dramatically reducing the angular momentum relaxation time \citep{chen&amaro-seoane2014}.
    \item There is a slight tension with the observed semimajor axis distribution of fainter S stars ($K < 16$), potentially indicating the progenitor binary population is more skewed to small separations than in \citet{moe&distefano2017}.
    \item The model KLF is in tension with observations, particularly with a Kroupa present-day mass function and realistic delay times between star formation and disruption. Potentially, this tension may be resolved with a top-heavy initial mass function or with a bursty star formation rate that is skewed toward the present day. However, reproducing the observed thermal eccentricity distribution would be more challenging in such scenarios.
    \item There is also tension with the measured ages of the bright S stars, with the model predicting older ages. Again, this may point to a bursty star formation rate. Alternatively, the measured ages may be underestimated if the S stars have subsolar metallicities.
\end{enumerate}


\begin{acknowledgements}
We thank the anonymous referee for helpful comments.
We are very grateful to our funding agencies.
AG was supported at the Technion by a Zuckerman Fellowship.
RS acknowledges financial support from the Severo Ochoa grant CEX2021-001131-S funded by MCIN/AEI/10.13039/501100011033, grant EUR2022-134031 funded by MCIN/AEI/10.13039/501100011033 and by the European Union NextGenerationEU/PRTR, and grant PID2022-136640NB-C21 funded by MCIN/AEI 10.13039/501100011033 and by the European Union. We thank Pau Amaro Seoane, Andreas Burkert, Sebastiano von Fellenberg, Jean-Baptiste Fouvry, Karamveer Kaur, Fabrice Martins, Nicholas Stone, Odelia Teboul, and Kerwann Tep for helpful comments and discussions. 
This work made use of the following software packages: Astropy \citep{astropy+2013,astropy+2018,astropy+2022}, Fewbody \citep{fregeau+2004}, JuDOKA \citep{tep+2021}, Jupyter \citep{kluyver+2016Jup}, Matplotlib \citep{hunter+2007}, MIST \citep{paxton+2011, paxton+2013, paxton+2015, dotter2016, choi+2016}, NumPy \citep{harris2020array}, SciPy \citep{2020SciPy-NMeth}, PARSEC \citep{bressan+2012, chen+2014, chen+2015, tang+2014, marigo+2017, pastorelli+2019, pastorelli+2020}, seaborn \citep{Waskom2021}.
\end{acknowledgements}

%

\vspace{5mm}




\bibliography{aanda}{}

\begin{thebibliography}{137}
\expandafter\ifx\csname natexlab\endcsname\relax\def\natexlab#1{#1}\fi

\bibitem[{{Aharon} \& {Perets}(2015)}]{aharon&perets2015}
{Aharon}, D. \& {Perets}, H.~B. 2015, \apj, 799, 185

\bibitem[{{Akiba} {et~al.}(2024){Akiba}, {Naoz}, \& {Madigan}}]{akiba+2024}
{Akiba}, T., {Naoz}, S., \& {Madigan}, A.-M. 2024, ApJL submitted,
  arXiv:2410.19881

\bibitem[{{Alexander}(2017)}]{alexander2017}
{Alexander}, T. 2017, ArXiv e-prints [\eprint[arXiv]{1701.04762}]

\bibitem[{{Alexander} \& {Hopman}(2009)}]{alexander&hopman2009}
{Alexander}, T. \& {Hopman}, C. 2009, \apj, 697, 1861

\bibitem[{{Alexander} \& {Morris}(2003)}]{alexander&morris2003}
{Alexander}, T. \& {Morris}, M. 2003, \apjl, 590, L25

\bibitem[{{Alexander} \& {Pfuhl}(2014)}]{alexander&pfuhl2014}
{Alexander}, T. \& {Pfuhl}, O. 2014, \apj, 780, 148

\bibitem[{{Amaro Seoane}(2023)}]{amaro-seoane2023}
{Amaro Seoane}, P. 2023, \apj, 947, 8

\bibitem[{{Antonini} {et~al.}(2010){Antonini}, {Faber}, {Gualandris}, \&
  {Merritt}}]{antonini+2010}
{Antonini}, F., {Faber}, J., {Gualandris}, A., \& {Merritt}, D. 2010, \apj,
  713, 90

\bibitem[{{Antonini} \& {Merritt}(2013)}]{antonini&merritt2013}
{Antonini}, F. \& {Merritt}, D. 2013, \apjl, 763, L10

\bibitem[{{Astropy Collaboration} {et~al.}(2022){Astropy Collaboration},
  {Price-Whelan}, {Lim}, {Earl}, {Starkman}, {Bradley}, {Shupe}, {Patil},
  {Corrales}, {Brasseur}, {N{\"o}the}, {Donath}, {Tollerud}, {Morris},
  {Ginsburg}, {Vaher}, {Weaver}, {Tocknell}, {Jamieson}, {van Kerkwijk},
  {Robitaille}, {Merry}, {Bachetti}, {G{\"u}nther}, {Aldcroft},
  {Alvarado-Montes}, {Archibald}, {B{\'o}di}, {Bapat}, {Barentsen},
  {Baz{\'a}n}, {Biswas}, {Boquien}, {Burke}, {Cara}, {Cara}, {Conroy},
  {Conseil}, {Craig}, {Cross}, {Cruz}, {D'Eugenio}, {Dencheva}, {Devillepoix},
  {Dietrich}, {Eigenbrot}, {Erben}, {Ferreira}, {Foreman-Mackey}, {Fox},
  {Freij}, {Garg}, {Geda}, {Glattly}, {Gondhalekar}, {Gordon}, {Grant},
  {Greenfield}, {Groener}, {Guest}, {Gurovich}, {Handberg}, {Hart},
  {Hatfield-Dodds}, {Homeier}, {Hosseinzadeh}, {Jenness}, {Jones}, {Joseph},
  {Kalmbach}, {Karamehmetoglu}, {Ka{\l}uszy{\'n}ski}, {Kelley}, {Kern},
  {Kerzendorf}, {Koch}, {Kulumani}, {Lee}, {Ly}, {Ma}, {MacBride}, {Maljaars},
  {Muna}, {Murphy}, {Norman}, {O'Steen}, {Oman}, {Pacifici}, {Pascual},
  {Pascual-Granado}, {Patil}, {Perren}, {Pickering}, {Rastogi}, {Roulston},
  {Ryan}, {Rykoff}, {Sabater}, {Sakurikar}, {Salgado}, {Sanghi}, {Saunders},
  {Savchenko}, {Schwardt}, {Seifert-Eckert}, {Shih}, {Jain}, {Shukla}, {Sick},
  {Simpson}, {Singanamalla}, {Singer}, {Singhal}, {Sinha}, {Sip{\H{o}}cz},
  {Spitler}, {Stansby}, {Streicher}, {{\v{S}}umak}, {Swinbank}, {Taranu},
  {Tewary}, {Tremblay}, {de Val-Borro}, {Van Kooten}, {Vasovi{\'c}}, {Verma},
  {de Miranda Cardoso}, {Williams}, {Wilson}, {Winkel}, {Wood-Vasey}, {Xue},
  {Yoachim}, {Zhang}, {Zonca}, \& {Astropy Project
  Contributors}}]{astropy+2022}
{Astropy Collaboration}, {Price-Whelan}, A.~M., {Lim}, P.~L., {et~al.} 2022,
  \apj, 935, 167

\bibitem[{{Astropy Collaboration} {et~al.}(2013){Astropy Collaboration},
  {Robitaille}, {Tollerud}, {Greenfield}, {Droettboom}, {Bray}, {Aldcroft},
  {Davis}, {Ginsburg}, {Price-Whelan}, {Kerzendorf}, {Conley}, {Crighton},
  {Barbary}, {Muna}, {Ferguson}, {Grollier}, {Parikh}, {Nair}, {Unther},
  {Deil}, {Woillez}, {Conseil}, {Kramer}, {Turner}, {Singer}, {Fox}, {Weaver},
  {Zabalza}, {Edwards}, {Azalee Bostroem}, {Burke}, {Casey}, {Crawford},
  {Dencheva}, {Ely}, {Jenness}, {Labrie}, {Lim}, {Pierfederici}, {Pontzen},
  {Ptak}, {Refsdal}, {Servillat}, \& {Streicher}}]{astropy+2013}
{Astropy Collaboration}, {Robitaille}, T.~P., {Tollerud}, E.~J., {et~al.} 2013,
  \aap, 558, A33

\bibitem[{{Bahcall} \& {Wolf}(1976)}]{bahcall&wolf1976}
{Bahcall}, J.~N. \& {Wolf}, R.~A. 1976, \apj, 209, 214

\bibitem[{{Balberg} {et~al.}(2013){Balberg}, {Sari}, \& {Loeb}}]{balberg+2013}
{Balberg}, S., {Sari}, R., \& {Loeb}, A. 2013, \mnras, 434, L26

\bibitem[{{Bar-Or} \& {Alexander}(2014)}]{bar-or&alexander2014}
{Bar-Or}, B. \& {Alexander}, T. 2014, Classical and Quantum Gravity, 31, 244003

\bibitem[{{Bar-Or} \& {Alexander}(2016)}]{bar-or&alexander2016}
{Bar-Or}, B. \& {Alexander}, T. 2016, \apj, 820, 129

\bibitem[{{Bar-Or} \& {Fouvry}(2018)}]{bar-or.fouvry2018}
{Bar-Or}, B. \& {Fouvry}, J.-B. 2018, \apjl, 860, L23

\bibitem[{{Bartko} {et~al.}(2009){Bartko}, {Martins}, {Fritz}, {Genzel},
  {Levin}, {Perets}, {Paumard}, {Nayakshin}, {Gerhard}, {Alexander},
  {Dodds-Eden}, {Eisenhauer}, {Gillessen}, {Mascetti}, {Ott}, {Perrin},
  {Pfuhl}, {Reid}, {Rouan}, {Sternberg}, \& {Trippe}}]{bartko+2009}
{Bartko}, H., {Martins}, F., {Fritz}, T.~K., {et~al.} 2009, \apj, 697, 1741

\bibitem[{{Baumgardt} {et~al.}(2018){Baumgardt}, {Amaro-Seoane}, \&
  {Sch{\"o}del}}]{baumgardt+2018}
{Baumgardt}, H., {Amaro-Seoane}, P., \& {Sch{\"o}del}, R. 2018, \aap, 609, A28

\bibitem[{{Bonnell} \& {Rice}(2008)}]{bonnell&rice2008}
{Bonnell}, I.~A. \& {Rice}, W.~K.~M. 2008, Science, 321, 1060

\bibitem[{{Bradnick} {et~al.}(2017){Bradnick}, {Mandel}, \&
  {Levin}}]{bradnick+2017}
{Bradnick}, B., {Mandel}, I., \& {Levin}, Y. 2017, \mnras, 469, 2042

\bibitem[{{Bressan} {et~al.}(2012){Bressan}, {Marigo}, {Girardi}, {Salasnich},
  {Dal Cero}, {Rubele}, \& {Nanni}}]{bressan+2012}
{Bressan}, A., {Marigo}, P., {Girardi}, L., {et~al.} 2012, \mnras, 427, 127

\bibitem[{{Burkert} {et~al.}(2024){Burkert}, {Gillessen}, {Lin}, {Zheng},
  {Schoeller}, {Eisenhauer}, \& {Genzel}}]{burkert+2023}
{Burkert}, A., {Gillessen}, S., {Lin}, D.~N.~C., {et~al.} 2024, \apj, 962, 81

\bibitem[{{Cao} {et~al.}(2024){Cao}, {Liu}, {Li}, {Chen}, \& {Wang}}]{cao+2024}
{Cao}, C.~Y., {Liu}, F.~K., {Li}, S., {Chen}, X., \& {Wang}, K. 2024, arXiv
  e-prints, arXiv:2411.09278

\bibitem[{{Chen} \& {Amaro-Seoane}(2014)}]{chen&amaro-seoane2014}
{Chen}, X. \& {Amaro-Seoane}, P. 2014, \apjl, 786, L14

\bibitem[{Chen {et~al.}(2015)Chen, Bressan, Girardi, Marigo, Kong, \&
  Lanza}]{chen+2015}
Chen, Y., Bressan, A., Girardi, L., {et~al.} 2015, Monthly Notices of the Royal
  Astronomical Society, 452, 1068

\bibitem[{Chen {et~al.}(2014)Chen, Girardi, Bressan, Marigo, Barbieri, \&
  Kong}]{chen+2014}
Chen, Y., Girardi, L., Bressan, A., {et~al.} 2014, Monthly Notices of the Royal
  Astronomical Society, 444, 2525

\bibitem[{{Choi} {et~al.}(2016){Choi}, {Dotter}, {Conroy}, {Cantiello},
  {Paxton}, \& {Johnson}}]{choi+2016}
{Choi}, J., {Dotter}, A., {Conroy}, C., {et~al.} 2016, \apj, 823, 102

\bibitem[{Collaboration {et~al.}(2024)Collaboration, Dayem, Abuter, Aimar,
  Seoane, Amorim, Beck, Berger, Bonnet, Bourdarot, Brandner, Cardoso, Dolcetta,
  Clénet, Davies, de~Zeeuw, Drescher, Eckart, Eisenhauer, Feuchtgruber,
  Finger, Schreiber, Foschi, Gao, Garcia, Gendron, Genzel, Gillessen, Hartl,
  Haubois, Haussman, Heißel, Hennig, Hippler, Horrobin, Jochum, Jocou, Kaufer,
  Kervella, Lacour, Lapeyrère, Bouquin, Léna, Lutz, Mang, More, Ott, Paumard,
  Perraut, Perrin, Pfuhl, Rabien, Ribeiro, Bordoni, Scheithauer, Shangguan,
  Shimizu, Stadler, Straub, Straubmeier, Sturm, Tacconi, Urso, Vincent,
  Fellenberg, Widmann, Wieprecht, Woillez, \& Zhang}]{gravity+2024extendedmass}
Collaboration, T.~G., Dayem, K. A.~E., Abuter, R., {et~al.} 2024, Improving
  constraints on the extended mass distribution in the Galactic Center with
  stellar orbits

\bibitem[{{Dale} {et~al.}(2009){Dale}, {Davies}, {Church}, \&
  {Freitag}}]{dale+2009}
{Dale}, J.~E., {Davies}, M.~B., {Church}, R.~P., \& {Freitag}, M. 2009, \mnras,
  393, 1016

\bibitem[{{Davies} {et~al.}(1998){Davies}, {Blackwell}, {Bailey}, \&
  {Sigurdsson}}]{davies+1998}
{Davies}, M.~B., {Blackwell}, R., {Bailey}, V.~C., \& {Sigurdsson}, S. 1998,
  \mnras, 301, 745

\bibitem[{{Dessart} {et~al.}(2024){Dessart}, {Ryu}, {Amaro Seoane}, \&
  {Taylor}}]{dessart+2024}
{Dessart}, L., {Ryu}, T., {Amaro Seoane}, P., \& {Taylor}, A.~M. 2024, \aap,
  682, A58

\bibitem[{{Do} {et~al.}(2019){Do}, {Hees}, {Ghez}, {Martinez}, {Chu}, {Jia},
  {Sakai}, {Lu}, {Gautam}, {O'Neil}, {Becklin}, {Morris}, {Matthews},
  {Nishiyama}, {Campbell}, {Chappell}, {Chen}, {Ciurlo}, {Dehghanfar},
  {Gallego-Cano}, {Kerzendorf}, {Lyke}, {Naoz}, {Saida}, {Sch{\"o}del},
  {Takahashi}, {Takamori}, {Witzel}, \& {Wizinowich}}]{do+2019}
{Do}, T., {Hees}, A., {Ghez}, A., {et~al.} 2019, Science, 365, 664

\bibitem[{{Do} {et~al.}(2015){Do}, {Kerzendorf}, {Winsor}, {St{\o}stad},
  {Morris}, {Lu}, \& {Ghez}}]{do+2015}
{Do}, T., {Kerzendorf}, W., {Winsor}, N., {et~al.} 2015, \apj, 809, 143

\bibitem[{Do {et~al.}(2013)Do, Lu, Ghez, Morris, Yelda, Martinez, Wright, \&
  Matthews}]{do+2013}
Do, T., Lu, J.~R., Ghez, A.~M., {et~al.} 2013, \apj, 764, 154

\bibitem[{{Dotter}(2016)}]{dotter2016}
{Dotter}, A. 2016, \apjs, 222, 8

\bibitem[{{Duncan} \& {Shapiro}(1983)}]{duncan&shapiro1983}
{Duncan}, M.~J. \& {Shapiro}, S.~L. 1983, \apj, 268, 565

\bibitem[{{Feldmeier-Krause} {et~al.}(2017){Feldmeier-Krause}, {Kerzendorf},
  {Neumayer}, {Sch{\"o}del}, {Nogueras-Lara}, {Do}, {de Zeeuw}, \&
  {Kuntschner}}]{feldmeier+2017Metal}
{Feldmeier-Krause}, A., {Kerzendorf}, W., {Neumayer}, N., {et~al.} 2017,
  \mnras, 464, 194

\bibitem[{{Fouvry} {et~al.}(2017){Fouvry}, {Pichon}, \&
  {Magorrian}}]{fouvry+2017}
{Fouvry}, J.~B., {Pichon}, C., \& {Magorrian}, J. 2017, \aap, 598, A71

\bibitem[{{Fragione} {et~al.}(2017){Fragione}, {Capuzzo-Dolcetta}, \&
  {Kroupa}}]{fragione+2017}
{Fragione}, G., {Capuzzo-Dolcetta}, R., \& {Kroupa}, P. 2017, \mnras, 467, 451

\bibitem[{{Fragione} \& {Sari}(2018)}]{fragione&sari2018}
{Fragione}, G. \& {Sari}, R. 2018, \apj, 852, 51

\bibitem[{{Fregeau} {et~al.}(2004){Fregeau}, {Cheung}, {Portegies Zwart}, \&
  {Rasio}}]{fregeau+2004}
{Fregeau}, J.~M., {Cheung}, P., {Portegies Zwart}, S.~F., \& {Rasio}, F.~A.
  2004, \mnras, 352, 1

\bibitem[{{Freitag} {et~al.}(2006{\natexlab{a}}){Freitag}, {Amaro-Seoane}, \&
  {Kalogera}}]{freitag+2006}
{Freitag}, M., {Amaro-Seoane}, P., \& {Kalogera}, V. 2006{\natexlab{a}}, \apj,
  649, 91

\bibitem[{{Freitag} {et~al.}(2006{\natexlab{b}}){Freitag}, {G{\"u}rkan}, \&
  {Rasio}}]{freitag+2006Coll}
{Freitag}, M., {G{\"u}rkan}, M.~A., \& {Rasio}, F.~A. 2006{\natexlab{b}},
  \mnras, 368, 141

\bibitem[{{Fritz} {et~al.}(2011){Fritz}, {Gillessen}, {Dodds-Eden}, {Lutz},
  {Genzel}, {Raab}, {Ott}, {Pfuhl}, {Eisenhauer}, \&
  {Yusef-Zadeh}}]{fritz+2011}
{Fritz}, T.~K., {Gillessen}, S., {Dodds-Eden}, K., {et~al.} 2011, \apj, 737, 73

\bibitem[{{Gallego-Cano} {et~al.}(2024){Gallego-Cano}, {Fritz}, {Sch{\"o}del},
  {Feldmeier-Krause}, {Do}, \& {Nishiyama}}]{gallego-cano+2024}
{Gallego-Cano}, E., {Fritz}, T., {Sch{\"o}del}, R., {et~al.} 2024, \aap, 689,
  A190

\bibitem[{{Gallego-Cano} {et~al.}(2018){Gallego-Cano}, {Sch{\"o}del}, {Dong},
  {Nogueras-Lara}, {Gallego-Calvente}, {Amaro-Seoane}, \&
  {Baumgardt}}]{gallego-cano+2018}
{Gallego-Cano}, E., {Sch{\"o}del}, R., {Dong}, H., {et~al.} 2018, \aap, 609,
  A26

\bibitem[{{Generozov}(2021)}]{generozov2021}
{Generozov}, A. 2021, \mnras, 501, 3088

\bibitem[{{Generozov} \& {Madigan}(2020)}]{generozov&madigan2020}
{Generozov}, A. \& {Madigan}, A.-M. 2020, \apj, 896, 137

\bibitem[{{Generozov} {et~al.}(2022){Generozov}, {Nayakshin}, \&
  {Madigan}}]{gnm+2022}
{Generozov}, A., {Nayakshin}, S., \& {Madigan}, A.~M. 2022, \mnras, 512, 4100

\bibitem[{{Generozov} {et~al.}(2018){Generozov}, {Stone}, {Metzger}, \&
  {Ostriker}}]{generozov+2018}
{Generozov}, A., {Stone}, N.~C., {Metzger}, B.~D., \& {Ostriker}, J.~P. 2018,
  \mnras, 478, 4030

\bibitem[{{Genzel} {et~al.}(1997){Genzel}, {Eckart}, {Ott}, \&
  {Eisenhauer}}]{genzel+1997}
{Genzel}, R., {Eckart}, A., {Ott}, T., \& {Eisenhauer}, F. 1997, \mnras, 291,
  219

\bibitem[{{Genzel} {et~al.}(1996){Genzel}, {Thatte}, {Krabbe}, {Kroker}, \&
  {Tacconi-Garman}}]{genzel+1996}
{Genzel}, R., {Thatte}, N., {Krabbe}, A., {Kroker}, H., \& {Tacconi-Garman},
  L.~E. 1996, \apj, 472, 153

\bibitem[{{Ghez} {et~al.}(2003){Ghez}, {Duch{\^e}ne}, {Matthews}, {Hornstein},
  {Tanner}, {Larkin}, {Morris}, {Becklin}, {Salim}, {Kremenek}, {Thompson},
  {Soifer}, {Neugebauer}, \& {McLean}}]{ghez+2003}
{Ghez}, A.~M., {Duch{\^e}ne}, G., {Matthews}, K., {et~al.} 2003, \apjl, 586,
  L127

\bibitem[{{Ghez} {et~al.}(1998){Ghez}, {Klein}, {Morris}, \&
  {Becklin}}]{ghez+1998}
{Ghez}, A.~M., {Klein}, B.~L., {Morris}, M., \& {Becklin}, E.~E. 1998, \apj,
  509, 678

\bibitem[{{Gillessen} {et~al.}(2017){Gillessen}, {Plewa}, {Eisenhauer}, {Sari},
  {Waisberg}, {Habibi}, {Pfuhl}, {George}, {Dexter}, \& {von
  Fellenberg}}]{gillessen+2017}
{Gillessen}, S., {Plewa}, P.~M., {Eisenhauer}, F., {et~al.} 2017, \apj, 837, 30

\bibitem[{{Ginsburg} \& {Loeb}(2007)}]{ginsburg&loeb2007}
{Ginsburg}, I. \& {Loeb}, A. 2007, \mnras, 376, 492

\bibitem[{{Gould} \& {Quillen}(2003)}]{Gou+03}
{Gould}, A. \& {Quillen}, A.~C. 2003, \apj, 592, 935

\bibitem[{{GRAVITY Collaboration} {et~al.}(2022){GRAVITY Collaboration},
  {Abuter}, {Aimar}, {Amorim}, {Ball}, {Baub{\"o}ck}, {Berger}, {Bonnet},
  {Bourdarot}, {Brandner}, {Cardoso}, {Cl{\'e}net}, {Dallilar}, {Davies}, {de
  Zeeuw}, {Dexter}, {Drescher}, {Eisenhauer}, {F{\"o}rster Schreiber},
  {Foschi}, {Garcia}, {Gao}, {Gendron}, {Genzel}, {Gillessen}, {Habibi},
  {Haubois}, {Hei{\ss}el}, {Henning}, {Hippler}, {Horrobin}, {Jochum}, {Jocou},
  {Kaufer}, {Kervella}, {Lacour}, {Lapeyr{\`e}re}, {Le Bouquin}, {L{\'e}na},
  {Lutz}, {Ott}, {Paumard}, {Perraut}, {Perrin}, {Pfuhl}, {Rabien},
  {Shangguan}, {Shimizu}, {Scheithauer}, {Stadler}, {Stephens}, {Straub},
  {Straubmeier}, {Sturm}, {Tacconi}, {Tristram}, {Vincent}, {von Fellenberg},
  {Widmann}, {Wieprecht}, {Wiezorrek}, {Woillez}, {Yazici}, \&
  {Young}}]{gravity+2022_mass}
{GRAVITY Collaboration}, {Abuter}, R., {Aimar}, N., {et~al.} 2022, \aap, 657,
  L12

\bibitem[{{GRAVITY Collaboration} {et~al.}(2018){GRAVITY Collaboration},
  {Abuter}, {Amorim}, {Anugu}, {Baub{\"o}ck}, {Benisty}, {Berger}, {Blind},
  {Bonnet}, {Brandner}, {Buron}, {Collin}, {Chapron}, {Cl{\'e}net}, {Coud{\'e}
  Du Foresto}, {de Zeeuw}, {Deen}, {Delplancke-Str{\"o}bele}, {Dembet},
  {Dexter}, {Duvert}, {Eckart}, {Eisenhauer}, {Finger}, {F{\"o}rster
  Schreiber}, {F{\'e}dou}, {Garcia}, {Garcia Lopez}, {Gao}, {Gendron},
  {Genzel}, {Gillessen}, {Gordo}, {Habibi}, {Haubois}, {Haug}, {Hau{\ss}mann},
  {Henning}, {Hippler}, {Horrobin}, {Hubert}, {Hubin}, {Jimenez Rosales},
  {Jochum}, {Jocou}, {Kaufer}, {Kellner}, {Kendrew}, {Kervella}, {Kok},
  {Kulas}, {Lacour}, {Lapeyr{\`e}re}, {Lazareff}, {Le Bouquin}, {L{\'e}na},
  {Lippa}, {Lenzen}, {M{\'e}rand}, {M{\"u}ler}, {Neumann}, {Ott}, {Palanca},
  {Paumard}, {Pasquini}, {Perraut}, {Perrin}, {Pfuhl}, {Plewa}, {Rabien},
  {Ram{\'\i}rez}, {Ramos}, {Rau}, {Rodr{\'\i}guez-Coira}, {Rohloff}, {Rousset},
  {Sanchez-Bermudez}, {Scheithauer}, {Sch{\"o}ller}, {Schuler}, {Spyromilio},
  {Straub}, {Straubmeier}, {Sturm}, {Tacconi}, {Tristram}, {Vincent}, {von
  Fellenberg}, {Wank}, {Waisberg}, {Widmann}, {Wieprecht}, {Wiest},
  {Wiezorrek}, {Woillez}, {Yazici}, {Ziegler}, \&
  {Zins}}]{gravity+2018_redshift}
{GRAVITY Collaboration}, {Abuter}, R., {Amorim}, A., {et~al.} 2018, \aap, 615,
  L15

\bibitem[{{GRAVITY Collaboration} {et~al.}(2020){GRAVITY Collaboration},
  {Abuter}, {Amorim}, {Baub{\"o}ck}, {Berger}, {Bonnet}, {Brandner}, {Cardoso},
  {Cl{\'e}net}, {de Zeeuw}, {Dexter}, {Eckart}, {Eisenhauer}, {F{\"o}rster
  Schreiber}, {Garcia}, {Gao}, {Gendron}, {Genzel}, {Gillessen}, {Habibi},
  {Haubois}, {Henning}, {Hippler}, {Horrobin}, {Jim{\'e}nez-Rosales}, {Jochum},
  {Jocou}, {Kaufer}, {Kervella}, {Lacour}, {Lapeyr{\`e}re}, {Le Bouquin},
  {L{\'e}na}, {Nowak}, {Ott}, {Paumard}, {Perraut}, {Perrin}, {Pfuhl},
  {Rodr{\'\i}guez-Coira}, {Shangguan}, {Scheithauer}, {Stadler}, {Straub},
  {Straubmeier}, {Sturm}, {Tacconi}, {Vincent}, {von Fellenberg}, {Waisberg},
  {Widmann}, {Wieprecht}, {Wiezorrek}, {Woillez}, {Yazici}, \&
  {Zins}}]{gravity+2020_S2prec}
{GRAVITY Collaboration}, {Abuter}, R., {Amorim}, A., {et~al.} 2020, \aap, 636,
  L5

\bibitem[{{GRAVITY Collaboration} {et~al.}(2021){GRAVITY Collaboration},
  {Abuter}, {Amorim}, {Baub{\"o}ck}, {Berger}, {Bonnet}, {Brandner},
  {Cl{\'e}net}, {Davies}, {de Zeeuw}, {Dexter}, {Dallilar}, {Drescher},
  {Eckart}, {Eisenhauer}, {F{\"o}rster Schreiber}, {Garcia}, {Gao}, {Gendron},
  {Genzel}, {Gillessen}, {Habibi}, {Haubois}, {Hei{\ss}el}, {Henning},
  {Hippler}, {Horrobin}, {Jim{\'e}nez-Rosales}, {Jochum}, {Jocou}, {Kaufer},
  {Kervella}, {Lacour}, {Lapeyr{\`e}re}, {Le Bouquin}, {L{\'e}na}, {Lutz},
  {Nowak}, {Ott}, {Paumard}, {Perraut}, {Perrin}, {Pfuhl}, {Rabien},
  {Rodr{\'\i}guez-Coira}, {Shangguan}, {Shimizu}, {Scheithauer}, {Stadler},
  {Straub}, {Straubmeier}, {Sturm}, {Tacconi}, {Vincent}, {von Fellenberg},
  {Waisberg}, {Widmann}, {Wieprecht}, {Wiezorrek}, {Woillez}, {Yazici},
  {Young}, \& {Zins}}]{gravity+2021_dist}
{GRAVITY Collaboration}, {Abuter}, R., {Amorim}, A., {et~al.} 2021, \aap, 647,
  A59

\bibitem[{{Gualandris} {et~al.}(2012){Gualandris}, {Mapelli}, \&
  {Perets}}]{gualandris+2012}
{Gualandris}, A., {Mapelli}, M., \& {Perets}, H.~B. 2012, \mnras, 427, 1793

\bibitem[{{Guillochon} \& {Ramirez-Ruiz}(2013)}]{guillochon&ramirezruiz2013}
{Guillochon}, J. \& {Ramirez-Ruiz}, E. 2013, \apj, 767, 25

\bibitem[{{G{\"u}rkan} {et~al.}(2004){G{\"u}rkan}, {Freitag}, \&
  {Rasio}}]{gurkan+2004}
{G{\"u}rkan}, M.~A., {Freitag}, M., \& {Rasio}, F.~A. 2004, \apj, 604, 632

\bibitem[{{Habibi} {et~al.}(2017){Habibi}, {Gillessen}, {Martins},
  {Eisenhauer}, {Plewa}, {Pfuhl}, {George}, {Dexter}, {Waisberg}, {Ott}, {von
  Fellenberg}, {Baub{\"o}ck}, {Jimenez-Rosales}, \& {Genzel}}]{habibi+2017}
{Habibi}, M., {Gillessen}, S., {Martins}, F., {et~al.} 2017, \apj, 847, 120

\bibitem[{Hailey {et~al.}(2018)Hailey, Mori, Bauer, Berkowitz, Hong, \&
  Hord}]{hailey+2018}
Hailey, C.~J., Mori, K., Bauer, F.~E., {et~al.} 2018, Nature, 556, 70

\bibitem[{{Hamers} \& {Perets}(2017)}]{Ham+17}
{Hamers}, A.~S. \& {Perets}, H.~B. 2017, \apj, 846, 123

\bibitem[{Harris {et~al.}(2020)Harris, Millman, van~der Walt, Gommers,
  Virtanen, Cournapeau, Wieser, Taylor, Berg, Smith, Kern, Picus, Hoyer, van
  Kerkwijk, Brett, Haldane, del R{\'{i}}o, Wiebe, Peterson,
  G{\'{e}}rard-Marchant, Sheppard, Reddy, Weckesser, Abbasi, Gohlke, \&
  Oliphant}]{harris2020array}
Harris, C.~R., Millman, K.~J., van~der Walt, S.~J., {et~al.} 2020, Nature, 585,
  357

\bibitem[{{Heggie}(1975)}]{heggie1975}
{Heggie}, D.~C. 1975, \mnras, 173, 729

\bibitem[{{Hills}(1983)}]{hills1983}
{Hills}, J.~G. 1983, \aj, 88, 1269

\bibitem[{{Hills}(1988)}]{hills1988}
{Hills}, J.~G. 1988, \nat, 331, 687

\bibitem[{{Hills}(1991)}]{hills1991}
{Hills}, J.~G. 1991, \aj, 102, 704

\bibitem[{{Hopman} \& {Alexander}(2006{\natexlab{a}})}]{hopman&alexander2006}
{Hopman}, C. \& {Alexander}, T. 2006{\natexlab{a}}, \apj, 645, 1152

\bibitem[{{Hopman} \& {Alexander}(2006{\natexlab{b}})}]{hopman&alexander2006a}
{Hopman}, C. \& {Alexander}, T. 2006{\natexlab{b}}, \apjl, 645, L133

\bibitem[{Hunter(2007)}]{hunter+2007}
Hunter, J.~D. 2007, Computing in Science \& Engineering, 9, 90

\bibitem[{{Kim} {et~al.}(2004){Kim}, {Figer}, \& {Morris}}]{kimfiger+2004}
{Kim}, S.~S., {Figer}, D.~F., \& {Morris}, M. 2004, \apjl, 607, L123

\bibitem[{Kluyver {et~al.}(2016)Kluyver, Ragan-Kelley, P{\'e}rez, Granger,
  Bussonnier, Frederic, Kelley, Hamrick, Grout, Corlay, Ivanov, Avila, Abdalla,
  \& Willing}]{kluyver+2016Jup}
Kluyver, T., Ragan-Kelley, B., P{\'e}rez, F., {et~al.} 2016, in Positioning and
  Power in Academic Publishing: Players, Agents and Agendas, ed. F.~Loizides \&
  B.~Schmidt, IOS Press, 87 -- 90

\bibitem[{{Kocsis} \& {Tremaine}(2011)}]{kocsis&tremaine2011}
{Kocsis}, B. \& {Tremaine}, S. 2011, \mnras, 412, 187

\bibitem[{{Kroupa}(2001)}]{kroupa2001}
{Kroupa}, P. 2001, \mnras, 322, 231

\bibitem[{{Levin} \& {Beloborodov}(2003)}]{levin&beloborodov2003}
{Levin}, Y. \& {Beloborodov}, A.~M. 2003, \apjl, 590, L33

\bibitem[{{Li} \& {Loeb}(2013)}]{li&loeb2013}
{Li}, G. \& {Loeb}, A. 2013, \mnras, 429, 3040

\bibitem[{{L{\"o}ckmann} {et~al.}(2008){L{\"o}ckmann}, {Baumgardt}, \&
  {Kroupa}}]{lockmann+2008}
{L{\"o}ckmann}, U., {Baumgardt}, H., \& {Kroupa}, P. 2008, \apjl, 683, L151

\bibitem[{{Lu} {et~al.}(2013){Lu}, {Do}, {Ghez}, {Morris}, {Yelda}, \&
  {Matthews}}]{lu+2013}
{Lu}, J.~R., {Do}, T., {Ghez}, A.~M., {et~al.} 2013, \apj, 764, 155

\bibitem[{{Lu} {et~al.}(2021){Lu}, {Fuller}, {Raveh}, {Perets}, {Li}, {Hosek},
  \& {Do}}]{wenbinlu+2020}
{Lu}, W., {Fuller}, J., {Raveh}, Y., {et~al.} 2021, \mnras, 503, 603

\bibitem[{{Maccarone} {et~al.}(2022){Maccarone}, {Degenaar}, {Tetarenko},
  {Heinke}, {Wijnands}, \& {Sivakoff}}]{maccarone+2022}
{Maccarone}, T.~J., {Degenaar}, N., {Tetarenko}, B.~E., {et~al.} 2022, \mnras,
  512, 2365

\bibitem[{Madigan {et~al.}(2009)Madigan, Levin, \& Hopman}]{madigan+2009}
Madigan, A.-M., Levin, Y., \& Hopman, C. 2009, \apjl, 697, L44

\bibitem[{{Mainetti} {et~al.}(2017){Mainetti}, {Lupi}, {Campana}, {Colpi},
  {Coughlin}, {Guillochon}, \& {Ramirez-Ruiz}}]{mainetti+2017}
{Mainetti}, D., {Lupi}, A., {Campana}, S., {et~al.} 2017, \aap, 600

\bibitem[{{Marigo} {et~al.}(2017){Marigo}, {Girardi}, {Bressan}, {Rosenfield},
  {Aringer}, {Chen}, {Dussin}, {Nanni}, {Pastorelli}, {Rodrigues}, {Trabucchi},
  {Bladh}, {Dalcanton}, {Groenewegen}, {Montalb{\'a}n}, \&
  {Wood}}]{marigo+2017}
{Marigo}, P., {Girardi}, L., {Bressan}, A., {et~al.} 2017, \apj, 835, 77

\bibitem[{{Merritt}(2013)}]{merritt2013}
{Merritt}, D. 2013, {Dynamics and Evolution of Galactic Nuclei} (Princeton
  University Press)

\bibitem[{{Merritt} {et~al.}(2011){Merritt}, {Alexander}, {Mikkola}, \&
  {Will}}]{merritt+2011}
{Merritt}, D., {Alexander}, T., {Mikkola}, S., \& {Will}, C.~M. 2011, \prd, 84,
  044024

\bibitem[{{Merritt} {et~al.}(2009){Merritt}, {Gualandris}, \&
  {Mikkola}}]{merritt+2009}
{Merritt}, D., {Gualandris}, A., \& {Mikkola}, S. 2009, \apjl, 693, L35

\bibitem[{{Moe} \& {Di Stefano}(2017)}]{moe&distefano2017}
{Moe}, M. \& {Di Stefano}, R. 2017, \apjs, 230, 15

\bibitem[{{Mori} {et~al.}(2019){Mori}, {Hailey}, {Mandel}, {Schutt},
  {Bachetti}, {Coerver}, {Baganoff}, {Dykaar}, {Grindlay}, {Haggard}, {Heuer},
  {Hong}, {Hord}, {Jin}, {Nynka}, {Ponti}, \& {Tomsick}}]{mori+2019}
{Mori}, K., {Hailey}, C.~J., {Mandel}, S., {et~al.} 2019, \apj, 885, 142

\bibitem[{{Mori} {et~al.}(2021){Mori}, {Hailey}, {Schutt}, {Mandel}, {Heuer},
  {Grindlay}, {Hong}, {Ponti}, \& {Tomsick}}]{mori+2021}
{Mori}, K., {Hailey}, C.~J., {Schutt}, T. Y.~E., {et~al.} 2021, \apj, 921, 148

\bibitem[{{Murphy} {et~al.}(1991){Murphy}, {Cohn}, \& {Durisen}}]{murphy+1991}
{Murphy}, B.~W., {Cohn}, H.~N., \& {Durisen}, R.~H. 1991, \apj, 370, 60

\bibitem[{{Nandakumar} {et~al.}(2018){Nandakumar}, {Ryde}, {Schultheis},
  {Thorsbro}, {J{\"o}nsson}, {Barklem}, {Rich}, \&
  {Fragkoudi}}]{nandakumar+2018}
{Nandakumar}, G., {Ryde}, N., {Schultheis}, M., {et~al.} 2018, \mnras, 478,
  4374

\bibitem[{{Nieuwmunster} {et~al.}(2024){Nieuwmunster}, {Schultheis}, {Sormani},
  {Fragkoudi}, {Nogueras-Lara}, {Sch{\"o}del}, {McMillan}, {Smith}, \&
  {Sanders}}]{nieuwmunster+2024}
{Nieuwmunster}, N., {Schultheis}, M., {Sormani}, M., {et~al.} 2024, \aap, 685,
  A93

\bibitem[{{Nogueras-Lara} {et~al.}(2024){Nogueras-Lara}, {Nieuwmunster},
  {Schultheis}, {Sormani}, {Fragkoudi}, {Thorsbro}, {Rich}, {Ryde}, {Sanders},
  \& {Smith}}]{NoguerasLara+2024}
{Nogueras-Lara}, F., {Nieuwmunster}, N., {Schultheis}, M., {et~al.} 2024, \aap,
  690, A313

\bibitem[{{Pastorelli} {et~al.}(2020){Pastorelli}, {Marigo}, {Girardi},
  {Aringer}, {Chen}, {Rubele}, {Trabucchi}, {Bladh}, {Boyer}, {Bressan},
  {Dalcanton}, {Groenewegen}, {Lebzelter}, {Mowlavi}, {Chubb}, {Cioni}, {de
  Grijs}, {Ivanov}, {Nanni}, {van Loon}, \& {Zaggia}}]{pastorelli+2020}
{Pastorelli}, G., {Marigo}, P., {Girardi}, L., {et~al.} 2020, \mnras, 498, 3283

\bibitem[{{Pastorelli} {et~al.}(2019){Pastorelli}, {Marigo}, {Girardi}, {Chen},
  {Rubele}, {Trabucchi}, {Aringer}, {Bladh}, {Bressan}, {Montalb{\'a}n},
  {Boyer}, {Dalcanton}, {Eriksson}, {Groenewegen}, {H{\"o}fner}, {Lebzelter},
  {Nanni}, {Rosenfield}, {Wood}, \& {Cioni}}]{pastorelli+2019}
{Pastorelli}, G., {Marigo}, P., {Girardi}, L., {et~al.} 2019, \mnras, 485, 5666

\bibitem[{{Paumard} {et~al.}(2006){Paumard}, {Genzel}, {Martins}, {Nayakshin},
  {Beloborodov}, {Levin}, {Trippe}, {Eisenhauer}, {Ott}, {Gillessen}, {Abuter},
  {Cuadra}, {Alexander}, \& {Sternberg}}]{paumard+2006}
{Paumard}, T., {Genzel}, R., {Martins}, F., {et~al.} 2006, \apj, 643, 1011

\bibitem[{{Paxton} {et~al.}(2011){Paxton}, {Bildsten}, {Dotter}, {Herwig},
  {Lesaffre}, \& {Timmes}}]{paxton+2011}
{Paxton}, B., {Bildsten}, L., {Dotter}, A., {et~al.} 2011, \apjs, 192, 3

\bibitem[{{Paxton} {et~al.}(2013){Paxton}, {Cantiello}, {Arras}, {Bildsten},
  {Brown}, {Dotter}, {Mankovich}, {Montgomery}, {Stello}, {Timmes}, \&
  {Townsend}}]{paxton+2013}
{Paxton}, B., {Cantiello}, M., {Arras}, P., {et~al.} 2013, \apjs, 208, 4

\bibitem[{{Paxton} {et~al.}(2015){Paxton}, {Marchant}, {Schwab}, {Bauer},
  {Bildsten}, {Cantiello}, {Dessart}, {Farmer}, {Hu}, {Langer}, {Townsend},
  {Townsley}, \& {Timmes}}]{paxton+2015}
{Paxton}, B., {Marchant}, P., {Schwab}, J., {et~al.} 2015, \apjs, 220, 15

\bibitem[{{Perets} \& {Gualandris}(2010)}]{perets.gualandris2010}
{Perets}, H.~B. \& {Gualandris}, A. 2010, \apj, 719, 220

\bibitem[{{Perets} {et~al.}(2009){Perets}, {Gualandris}, {Kupi}, {Merritt}, \&
  {Alexander}}]{perets+2009}
{Perets}, H.~B., {Gualandris}, A., {Kupi}, G., {Merritt}, D., \& {Alexander},
  T. 2009, \apj, 702, 884

\bibitem[{Perets {et~al.}(2007)Perets, Hopman, \& Alexander}]{perets+2007}
Perets, H.~B., Hopman, C., \& Alexander, T. 2007, \apj, 656, 709

\bibitem[{Peters(1964)}]{peters1964}
Peters, P.~C. 1964, Physical Review, 136, 1224

\bibitem[{{Portegies Zwart} {et~al.}(1999){Portegies Zwart}, {Makino},
  {McMillan}, \& {Hut}}]{portegies-zwart+1999}
{Portegies Zwart}, S.~F., {Makino}, J., {McMillan}, S.~L.~W., \& {Hut}, P.
  1999, \aap, 348, 117

\bibitem[{{Preto} \& {Amaro-Seoane}(2010)}]{preto&amaro-seoane2010}
{Preto}, M. \& {Amaro-Seoane}, P. 2010, \apjl, 708, L42

\bibitem[{Rantala \& Naab(2023)}]{rantala&naab2023}
Rantala, A. \& Naab, T. 2023, Monthly Notices of the Royal Astronomical
  Society, 527, 11458

\bibitem[{{Rauch} \& {Tremaine}(1996)}]{rauch&tremaine1996}
{Rauch}, K.~P. \& {Tremaine}, S. 1996, \na, 1, 149

\bibitem[{{Rees}(1988)}]{rees1988}
{Rees}, M.~J. 1988, \nat, 333, 523

\bibitem[{{Rich} {et~al.}(2017){Rich}, {Ryde}, {Thorsbro}, {Fritz},
  {Schultheis}, {Origlia}, \& {J{\"o}nsson}}]{rich+2017}
{Rich}, R.~M., {Ryde}, N., {Thorsbro}, B., {et~al.} 2017, \aj, 154, 239

\bibitem[{{Rose} {et~al.}(2023){Rose}, {Naoz}, {Sari}, \& {Linial}}]{rose+2023}
{Rose}, S.~C., {Naoz}, S., {Sari}, R., \& {Linial}, I. 2023, \apj, 955, 30

\bibitem[{{Ryu} {et~al.}(2024){Ryu}, {Amaro Seoane}, {Taylor}, \&
  {Ohlmann}}]{ryu+2024a}
{Ryu}, T., {Amaro Seoane}, P., {Taylor}, A.~M., \& {Ohlmann}, S.~T. 2024,
  \mnras, 528, 6193

\bibitem[{{Ryu} {et~al.}(2020{\natexlab{a}}){Ryu}, {Krolik}, {Piran}, \&
  {Noble}}]{ryu+2020}
{Ryu}, T., {Krolik}, J., {Piran}, T., \& {Noble}, S.~C. 2020{\natexlab{a}},
  \apj, 904, 98

\bibitem[{{Ryu} {et~al.}(2020{\natexlab{b}}){Ryu}, {Krolik}, {Piran}, \&
  {Noble}}]{ryu+2020part}
{Ryu}, T., {Krolik}, J., {Piran}, T., \& {Noble}, S.~C. 2020{\natexlab{b}},
  \apj, 904, 100

\bibitem[{{Sanders}(1992)}]{sanders1992}
{Sanders}, R.~H. 1992, \nat, 359, 131

\bibitem[{Sari {et~al.}(2010)Sari, Kobayashi, \& Rossi}]{sari+2010}
Sari, R., Kobayashi, S., \& Rossi, E.~M. 2010, \apj, 708, 605

\bibitem[{{Sch{\"o}del} {et~al.}(2018){Sch{\"o}del}, {Gallego-Cano}, {Dong},
  {Nogueras-Lara}, {Gallego-Calvente}, {Amaro-Seoane}, \&
  {Baumgardt}}]{schodel+2018}
{Sch{\"o}del}, R., {Gallego-Cano}, E., {Dong}, H., {et~al.} 2018, \aap, 609,
  A27

\bibitem[{{Sch{\"o}del} {et~al.}(2010){Sch{\"o}del}, {Najarro}, {Muzic}, \&
  {Eckart}}]{schodel+2010extinction}
{Sch{\"o}del}, R., {Najarro}, F., {Muzic}, K., \& {Eckart}, A. 2010, \aap, 511,
  A18

\bibitem[{{Sch{\"o}del} {et~al.}(2020){Sch{\"o}del}, {Nogueras-Lara},
  {Gallego-Cano}, {Shahzamanian}, {Gallego-Calvente}, \&
  {Gardini}}]{schodel+2020}
{Sch{\"o}del}, R., {Nogueras-Lara}, F., {Gallego-Cano}, E., {et~al.} 2020,
  \aap, 641, A102

\bibitem[{{Sch{\"o}del} {et~al.}(2023){Sch{\"o}del}, {Nogueras-Lara}, {Hosek},
  {Do}, {Lu}, {Mart{\'\i}nez Arranz}, {Ghez}, {Rich}, {Gardini},
  {Gallego-Cano}, {Cano Gonz{\'a}lez}, \& {Gallego-Calvente}}]{schodel+2023}
{Sch{\"o}del}, R., {Nogueras-Lara}, F., {Hosek}, M., {et~al.} 2023, \aap, 672,
  L8

\bibitem[{{Schultheis} {et~al.}(2021){Schultheis}, {Fritz}, {Nandakumar},
  {Rojas-Arriagada}, {Nogueras-Lara}, {Feldmeier-Krause}, {Gerhard},
  {Neumayer}, {Patrick}, {Prieto}, {Sch{\"o}del}, {Mastrobuono-Battisti}, \&
  {Sormani}}]{schultheis+2021}
{Schultheis}, M., {Fritz}, T.~K., {Nandakumar}, G., {et~al.} 2021, \aap, 650,
  A191

\bibitem[{{Schultheis} {et~al.}(2019){Schultheis}, {Rich}, {Origlia}, {Ryde},
  {Nandakumar}, {Thorsbro}, \& {Neumayer}}]{schultheis+2019}
{Schultheis}, M., {Rich}, R.~M., {Origlia}, L., {et~al.} 2019, \aap, 627, A152

\bibitem[{{Spitzer} \& {Saslaw}(1966)}]{spitzer&saslaw1966}
{Spitzer}, Lyman, J. \& {Saslaw}, W.~C. 1966, \apj, 143, 400

\bibitem[{{Sridhar} \& {Touma}(2016)}]{sridhar&touma2016}
{Sridhar}, S. \& {Touma}, J.~R. 2016, \mnras, 458, 4143

\bibitem[{Tang {et~al.}(2014)Tang, Bressan, Rosenfield, Slemer, Marigo,
  Girardi, \& Bianchi}]{tang+2014}
Tang, J., Bressan, A., Rosenfield, P., {et~al.} 2014, Monthly Notices of the
  Royal Astronomical Society, 445, 4287

\bibitem[{{Tep} {et~al.}(2021){Tep}, {Fouvry}, {Pichon}, {Hei{\ss}el},
  {Paumard}, {Perrin}, \& {Vincent}}]{tep+2021}
{Tep}, K., {Fouvry}, J.-B., {Pichon}, C., {et~al.} 2021, \mnras, 506, 4289

\bibitem[{{The Astropy Collaboration} {et~al.}(2018){The Astropy
  Collaboration}, {Price-Whelan}, {Sip{\H o}cz}, {G{\"u}nther}, {Lim},
  {Crawford}, \& {Contributors}}]{astropy+2018}
{The Astropy Collaboration}, {Price-Whelan}, A.~M., {Sip{\H o}cz}, B.~M.,
  {et~al.} 2018, \aj, 156, 123

\bibitem[{{Vasiliev}(2017)}]{vasiliev2017}
{Vasiliev}, E. 2017, \apj, 848, 10

\bibitem[{Virtanen {et~al.}(2020)Virtanen, Gommers, Oliphant, Haberland, Reddy,
  Cournapeau, Burovski, Peterson, Weckesser, Bright, van~der Walt, Brett,
  Wilson, Millman, Mayorov, Nelson, Jones, Kern, Larson, Carey, Polat, Feng,
  Moore, VanderPlas, Laxalde, Perktold, Cimrman, Henriksen, Quintero, Harris,
  Archibald, Ribeiro, Pedregosa, van Mulbregt, Vijaykumar, Bardelli, Rothberg,
  Hilboll, Kloeckner, Scopatz, Lee, Rokem, Woods, Fulton, Masson,
  H{\"a}ggstr{\"o}m, Fitzgerald, Nicholson, Hagen, Pasechnik, Olivetti, Martin,
  Wieser, Silva, Lenders, Wilhelm, Young, Price, Ingold, Allen, Lee, Audren,
  Probst, Dietrich, Silterra, Webber, Slavi{\v{c}}, Nothman, Buchner, Kulick,
  Sch{\"o}nberger, de~Miranda~Cardoso, Reimer, Harrington, Rodr{\'i}guez,
  Nunez-Iglesias, Kuczynski, Tritz, Thoma, Newville, K{\"u}mmerer, Bolingbroke,
  Tartre, Pak, Smith, Nowaczyk, Shebanov, Pavlyk, Brodtkorb, Lee, McGibbon,
  Feldbauer, Lewis, Tygier, Sievert, Vigna, Peterson, More, Pudlik, Oshima,
  Pingel, Robitaille, Spura, Jones, Cera, Leslie, Zito, Krauss, Upadhyay,
  Halchenko, V{\'a}zquez-Baeza, \& 1.0~Contributors}]{2020SciPy-NMeth}
Virtanen, P., Gommers, R., Oliphant, T.~E., {et~al.} 2020, Nature Methods, 17,
  261

\bibitem[{{von Fellenberg} {et~al.}(2022){von Fellenberg}, {Gillessen},
  {Stadler}, {Baub{\"o}ck}, {Genzel}, {de Zeeuw}, {Pfuhl}, {Amaro Seoane},
  {Drescher}, {Eisenhauer}, {Habibi}, {Ott}, {Widmann}, \&
  {Young}}]{vonFellenberg+2022}
{von Fellenberg}, S.~D., {Gillessen}, S., {Stadler}, J., {et~al.} 2022, \apjl,
  932, L6

\bibitem[{Waskom(2021)}]{Waskom2021}
Waskom, M.~L. 2021, Journal of Open Source Software, 6, 3021

\bibitem[{{Yelda} {et~al.}(2014){Yelda}, {Ghez}, {Lu}, {Do}, {Meyer}, {Morris},
  \& {Matthews}}]{yelda+2014}
{Yelda}, S., {Ghez}, A.~M., {Lu}, J.~R., {et~al.} 2014, \apj, 783, 131

\bibitem[{{Zhang} \& {Amaro Seoane}(2024)}]{zhang&amaro-seoane2024}
{Zhang}, F. \& {Amaro Seoane}, P. 2024, \apj, 961, 232

\end{thebibliography}
\bibliographystyle{aa}

\begin{appendix}

\onecolumn
\section{Binary evaporation}
\label{sec:evap}
In principle, the binary population in the Galactic center region may be shaped by dynamical processes. For example, soft binaries can be evaporated by encounters with other stars. Here we quantify this effect and show that in our 
models binaries are more likely to disrupt rather than evaporate, and thus binary evaporation may be neglected.

The local evaporation timescale is (see \citealp{alexander&pfuhl2014})

\begin{align}
t_{\rm evap} &= 0.07 \frac{v_{12}^2 \sigma}{G^2 n <m^2> \ln \Lambda_{12}}\nonumber \\
& = 2.6 \times 10^{10} {\rm yr} \left(\frac{a_{\rm bin}}{15 {\rm au}}\right)^{-1} \left(\frac{<m^2>}{1 M_{\odot}}\right)^{-1} \left(\frac{m_{\rm bin}}{6 M_{\odot}}\right) \nonumber\\ 
&\left(\frac{\sigma}{100 {\rm km s^{-1}}}\right) \left(\frac{n}{5000 {\rm pc^{-3}}}\right) \ln \Lambda_{12}^{-1},
\end{align}
where $v_{12}$ is the internal velocity of the binary, $n$, $\sigma$, and $<m^2>$ are the number density, velocity dispersion of the perturbing population, and $\Lambda_{12} = 2 \sigma^2 / v_{12}^2$.
In our model, binaries come from outside the central 5 pc, where $n\approx 5\times 10^3$ pc$^{-3}$, and $\sigma\approx 100$ km s$^{-1}$.
For binaries with semimajor axes $\leq 15$ au in this region, the evaporation
timescale is of order 10 Gyr or more.

In comparison, the timescale for binaries to disrupt is
\begin{equation}
t_{\rm dis} = \frac{r}{\sigma} \frac{j_c^2}{j_{\rm lc}^2},
\end{equation}
where $j_{lc}=\sqrt{2 G M_{\rm bh} r_t}$. For an enclosed mass of $10^7 M_{\odot}$ inside 5 pc, 

\begin{equation}
t_{\rm dis}(5 {\rm pc}) \approx 3\times 10^7 {\rm yr} \left(\frac{\sigma}{100 {\rm km s^{-1}}}\right)^{-1} \left(\frac{m_{\rm bin}}{6 M_{\odot}}\right)^{-1/3} \left(\frac{a_{\rm bin}}{15 {\rm au}}\right)^{-1} 
\end{equation}
Binaries on eccentric orbits will cross the central region, where the evaporation timescale is shorter. In particular, the evaporation timescale will be between $r^{1.3} - r^{1.5}$.\footnote{For black hole density profiles between $r^{-1.8}$ and $r^{-2}$.} The average evaporation time over the binary population is

\begin{equation}
<t_{\rm evap}> = \int_{0}^{e_{\rm max}} de f(e) \left(\int_{r_p}^{r_a} dr \frac{2}{P t_{\rm evap}} \frac{1}{v_r}\right)^{-1},
\end{equation}
where $r_p$, $r_a$, $v_r$, $P$, $e$, and $f(e)$ are the pericenter, apocenter, radial velocity, orbital period, eccentricity, and eccentricity distribution respectively. Note that the inner integral corresponds to a harmonic average. 

In steady state $f(e)$ is expected to be thermal. Also, since we are considering binaries that are not too far outside the SMBH sphere of influence, we assume Keplerian scalings with radius for all the dynamical quantities. 
Overall, we find eccentric orbits give a $\sim 10\%$ correction to the evaporation time.

Although we have focused on soft binaries, the hardening rate of hard binaries has a similar dependence on cluster and binary properties \citep{heggie1975, hills1983}. Therefore, we also expect hard binaries to disrupt before they can significantly harden.

\section{K-band luminosity function}
\label{sec:sKLF}
Here, we describe how we derive the KLF of early stars inside 0.5'' (the red line in Figure~\ref{fig:kmag}) from the data in \citet{schodel+2020}.

First, we compute the KLF in two radial rings with $r_{\rm proj}<0.5\arcsec$ and $4\arcsec<r_{\rm proj}<5\arcsec$.  These KLFs include contribution from both early and late stars, with surface density profiles  $r^{-0.8 \pm 0.08}$ and $r^{0\pm 0.08}$, respectively (Table 6 in \citealp{gallego-cano+2024}). The surface densities are equal at $\sim0.7\arcsec$, such that late (early) stars dominate on larger (smaller) scales.

The KLF of early stars within $0.5\arcsec$ is

\begin{align}
    K_{\rm early, 0.5}\approx K_{0.5} - W K_{4}
    \label{eq:k1}
\end{align}
where $K_{0.5}$ is the total KLF inside $0.5\arcsec$ and $K_{4}$ is the KLF between $4\arcsec$ and $5\arcsec$, which is dominated by late stars. The weight, W, is chosen to reproduce the ratio of late to all stars inside $0.5\arcsec$, $X$. From the surface density profiles in \citet{gallego-cano+2024}, $X\approx0.3$ and

\begin{align}
   W=\frac{N(r_{\rm proj}<0.5\arcsec, K \leq 16.15)} {N(4\arcsec<r_{\rm proj}<5\arcsec, K \leq 16.15)}  X \approx 1.3,
   \label{eq:k2}
\end{align}
where the numerator is the number of stars with projected radius, $r_{\rm proj}$ less than $0.5\arcsec$ and $K \leq 16$. Similarly, the denominator is the number of stars with $4\arcsec<r_{\rm proj}<5\arcsec$ and $K \leq 16.15$.
The cut near $K=16.15$ is motivated by exclusion of fainter stars from the surface density fits in \citet{gallego-cano+2024}. 

From equations~\eqref{eq:k1} and~\eqref{eq:k2}, we obtain the $K_{\rm early, 0.5}$, the red line in Figure~\ref{fig:kmag}. The uncertainty is calculated by adding the uncertainties from $K_{0.5}$ and $K_{4}$ in quadrature. 
The results change by less than the uncertainty, if we use the KLF from $2$--$3\arcsec$ or $3$--$4\arcsec$ in lieu of $K_4$. In fact, the difference between $K_{0.5}$ and $K_{\rm early, 0.5}$ are smaller than the uncertainties for $K \lesssim 19$.

\section{S-star ages versus metallicity}
\label{sec:metal}
\begin{figure}[h!]
    \centering
    \includegraphics[width=0.5\columnwidth]{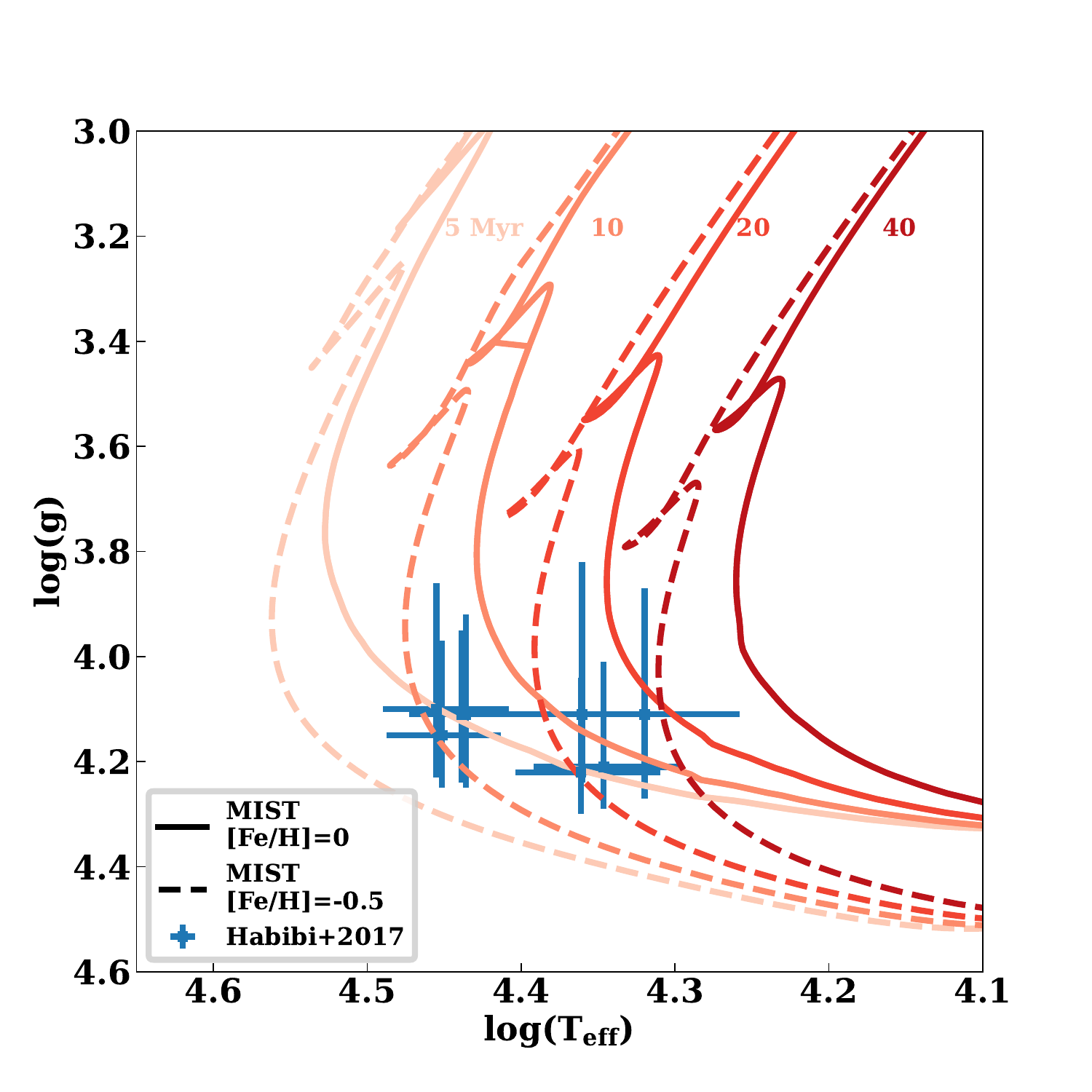}
    \caption{$\log(g)$ versus $\log(T_{\rm eff})$ for the eight bright S stars from \citet{habibi+2017} compared to MIST isochrones (v1.2, $v/v_{\rm crit}=0.4$). The solid (dashed) lines correspond to $[{\rm Fe/H}]=0$ ($[{\rm Fe/H}]=-0.5$).}
    \label{fig:ageMetal}
\end{figure}

Here we show that the S-star ages are degenerate with metallicity. Figure~\ref{fig:ageMetal} shows a comparison of MIST isochrones of $\log(g)$ versus $\log(T_{\rm eff})$ for two different metallicities. Qualitatively, the S stars can either be younger and metal-rich or older and metal-poor. For each star, we estimate the age by identifying the closest main-sequence isochrone point.\footnote{We minimize $(\log(L_{\rm obs}) -  \log(L_{\rm iso}))^2 + (\log(T_{\rm eff, obs}) -  \log(T_{\rm eff, iso}))^2 + (\log(g_{\rm obs}) -  \log(g_{\rm iso}))^2$, where the subscripts ``${\rm obs}$'' and ``${\rm iso}$'' indicate observed (from \citealt{habibi+2017}) and isochrone values respectively.} We summarize these estimates in Table~\ref{tab:sages}, along with the fits from \citet{habibi+2017}. We emphasize that our ages are merely estimates to illustrate the trend with metallicity, rather than detailed fits.

\begin{table}[h!]
\centering
\caption{Ages for eight bright S stars from \citet{habibi+2017}, and from MIST isochrones at different metallicities.}
\label{tab:sages}
\begin{tabular}{lcccc}
\hline
Star & Habibi et al. 2017 & MIST, [Fe/H]=0.0 & MIST, [Fe/H]=-0.5 & MIST, [Fe/H]=-1.0 \\
     & [Myr]     & [Myr]           & [Myr]            & [Myr]   \\
\hline
\hline
S1 & ${4.3}_{-4.2}^{+4.2}$ & 5.0 & 10.0 & 14.1 \\
S2 & ${6.6}_{-4.7}^{+3.4}$ & 4.5 & 7.9 & 11.2 \\
S4 & ${5.9}_{-5.5}^{+4.1}$ & 5.0 & 10.0 & 14.1 \\
S6 & ${11.5}_{-9.3}^{+7.8}$ & 8.9 & 17.8 & 25.1 \\
S7 & ${0.3}_{-0.3}^{+14.3}$ & 4.0 & 15.8 & 25.1 \\
S8 & ${3.1}_{-3.0}^{+4.4}$ & 3.2 & 7.9 & 11.2 \\
S9 & ${3.5}_{-3.5}^{+13.2}$ & 5.0 & 15.8 & 25.1 \\
S12 & ${14.7}_{-13.9}^{+10.6}$ & 14.1 & 28.2 & 35.5 \\
\hline
\end{tabular}
\end{table}


\end{appendix}
\label{LastPage}
\end{document}